\documentclass[aps,onecolumn,nofootinbib]{revtex4-1}
\usepackage{epsfig}
\usepackage{color}
\usepackage{hyperref}
\usepackage{mathbbol}
\usepackage{slashed}
\usepackage{amsmath}
\usepackage{amssymb}
\usepackage{diagbox}
\usepackage{xcolor,colortbl}

\definecolor{Gray}{gray}{0.8}
\definecolor{Grayy}{gray}{0.95}
\definecolor{LightCyan}{rgb}{0.88,1,1}
\newcolumntype{a}{>{\columncolor{Gray}}c}
\newcolumntype{b}{>{\columncolor{Grayy}}c}

\newcommand{\bs}[1]{\ensuremath{\boldsymbol{#1}}}
\newcommand{\eodim}{B_{d}^{}}
\newcommand{\eotet}{B_{\alpha}^{}}
\newcommand{\eotri}{B_{\text{\tiny t}}^{}}
\newcommand{\eonn}{B_{\text{\tiny nn}}^{}}

\newcommand{\mn}{m_\text{\tiny N}}
\newcommand{\mpi}{$m_\pi$}
\newcommand{\mpill}{\mbox{$m_\pi\sim 806~$MeV}}
\newcommand{\mpif}{\mbox{$m_\pi\sim 450~$MeV}}

\newcommand{\ie}{\textit{i.e.}}
\newcommand{\wrt}{\textit{w.r.t.}}
\newcommand{\eg}{\textit{e.g.}\;}

\newcommand{\anps}{{}^1a_{np}}
\newcommand{\anpt}{{}^3a_{np}}
\newcommand{\anpst}{{}^{1,3}a_{np}}

\newcommand{\andd}{{}^2a_{nd}}
\newcommand{\andq}{{}^4a_{nd}}

\newcommand{\eftnopi}{\mbox{EFT($\slashed{\pi}$)}}

\newcommand{\goodchi}{\protect\raisebox{2pt}{$\chi$}}
\newcommand{\ce}{\goodchi EFT}

\newcommand{\scite}[1]{\hspace*{-0.2em}$\textsuperscript{\cite{#1}}$}
\begin{document}
\markboth{Johannes Kirscher}{Matching effective few-nucleon theories to QCD}
\title{Matching effective few-nucleon theories to QCD}
\author{J.~Kirscher}
\address{Racah Institute of Physics, The Hebrew University, Jerusalem 91904, Israel}
\begin{abstract}
The emergence of complex macroscopic phenomena from a small set of
parameters and microscopic concepts demonstrates the power and
beauty of physical theories.
A theory which relates the wealth of data and peculiarities found in nuclei to the small
number of parameters and symmetries of quantum chromodynamics is by that standard of
exceptional beauty.

Decade-long research on computational physics and on effective field theories
facilitate the assessment of the presumption that quark masses and strong and
electromagnetic coupling constants suffice to parameterize the nuclear chart.
By presenting the current status of that enterprise, this article touches
the methodology of predicting nuclei  by simulating the constituting quarks and gluons
and the development of effective field theories as appropriate representations
of the fundamental theory.

While the nuclear spectra and electromagnetic responses analyzed computationally
so far with lattice QCD are in close resemblance to those which intrigued experimentalists a century ago,
they also test the theoretical understanding 
which was unavailable to guide the nuclear pioneers but developed since then. This
understanding is shown to be deficient in terms of correlations amongst
nuclear observables and their sensitivity to fundamental parameters.
By reviewing the transition from one effective field theory to another,
from QCD to pionful chiral theories to pionless and eventually to cluster
theories, we identify some of those deficiencies and conceptual problems
awaiting a solution before QCD can be identified as the high-energy theory
from which the nuclear landscape emerges.
\end{abstract}
\maketitle
\section{Introduction}
\paragraph*{Lattice QCD as an experiment with light baryons  ---}
It is a striking analogy that the calculation of binding energies of the lightest nuclei
as compound objects, whose constituent dynamics are dictated by quantum chromodynamics (QCD),
proceeds along similar paths as the early experiments in the real world.
The relative determination of the deuteron mass to Helium mass\scite{PhysRev.43.103},
with its preceding preparation of a sufficiently pure beam of the respective nuclei,
resembles remarkably the lattice methodology. In the latter, mass ratios are also more accessible
because of a reduced statistical uncertainty. Furthermore, devising interpolating operators
with a large initial overlap with the baryons of interest are the numerical analog of preparing
a clean beam.
The exploration of the nuclear landscape included in parallel
to the study of mass spectra the response of nuclei to electromagnetic fields\scite{PhysRev.56.728}.
Again, these investigations find their resemblance in contemporary
lattice calculations of the magnetic structure of light nuclei.
The fact that the lattice simulates nuclei at artificially large pion masses yields more
than just data points for an extrapolation to the physical pion mass.
It allows for the investigation of phenomena emergent from a theory as sophisticated as the one we
deem appropriate for the strong force in nature. The theory at unphysical pion masses shares
symmetries and degrees of freedom and thereby can identify features that are peculiar to QCD
as the theory which breaks the flavor symmetry with some specific quark-mass values \textit{or}
can be universally attributed to the three SU(3) color \textit{and} flavor symmetric quarks.

Tools that have evolved over the century and the questions which can be investigated with the lattice 
technology are plenty. The basic motivation, to build a theory
on the available data to predict what presumably will be measured with
more effort in nuclei with a total number of neutrons and protons $A\gtrsim4$, using different 
external probes, or looking at the systems at different scales,
remains, however, invariant. Today, we start at the same point like physicists
when facing the first measurements on light nuclei in the physical world.
\paragraph*{Contemporary theoretical nuclear reality ---}
With the increasing resolution of experimental apparatus and the ensuing discovery of the substructure of nucleons
the interaction amongst them lost its status of a fundamental theory and became to be thought of as a
manifestation of QCD at low energy. The current understanding of nuclei comprises a set of effective theories, each devised to
describe them only up to a certain nucleon number $A$ or momentum scale $Q$, at which the neutrons and protons 
can be excited and/or deformed but their internal structure is not probed.
The pioneering theories and models for the nuclear interaction, \eg, meson-exchange (Yukawa)
theories and quark models, can be recovered\footnote{See Ref.~\cite{Durand:2001zz} as an example for a derivation of a quark model from the heavy-baryon effective field theory of QCD.}
from or found to be included in these effective theories.
\begin{figure}[tb]
\centerline{\includegraphics{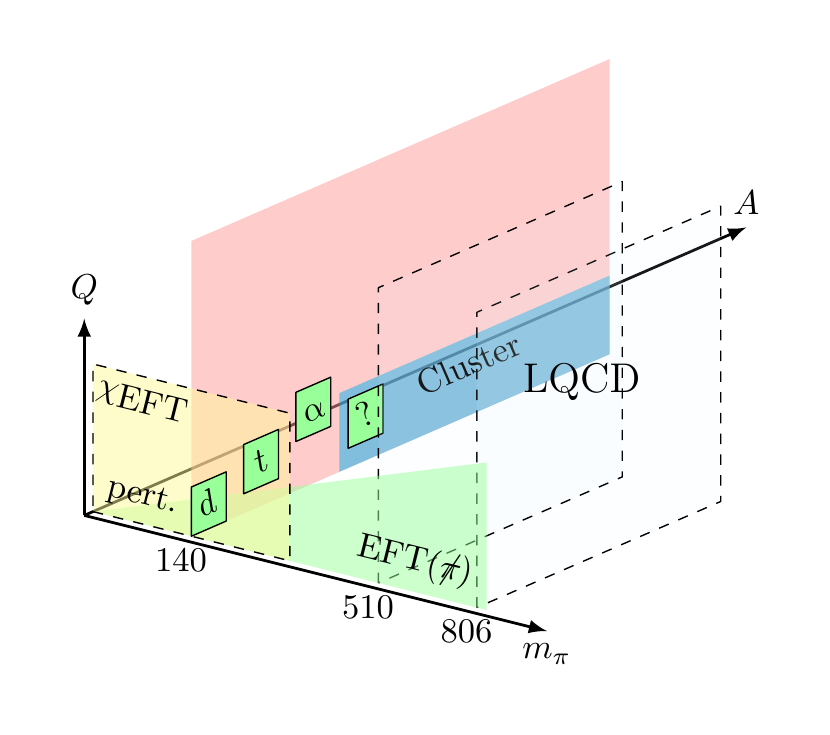}}
\caption{\label{fig.theoland} (Color online) Range of applicability of nuclear effective
theories (adapted from Ref.~\protect\cite{uvk-priv-comm}~)~
in a space spanned by the pion mass \mpi, a low-momentum scale $Q$, and the nucleon number $A$. The physical world is measured at $m_\pi\sim140~$MeV on the red sheet.
LQCD assesses the few-nucleon systems of the deuteron $d$, the triton $t$, and the $\alpha$-particle on the dashed enclosed sheets
at 510 and 806~MeV.
\eftnopi~(light green) was applied to $A<5$ nuclei up to large pion masses. \ce~(yellow area) breaks down at some
\mpi, but applies at larger momenta than its perturbative version and \eftnopi. Cluster EFTs (cyan) have only
been used at physical \mpi.}
\end{figure}

Specifically, we understand the nuclear interaction to be given by chiral
effective field theory (\ce) which:
\textit{(i)} treats the pion-nucleon coupling perturbatively for very low momenta,
\textit{(ii)} becomes non-perturbative for higher energies,
\mbox{\textit{(iii)} can} be matched to the pionless effective field theory (\eftnopi)
which is applicable in the baryon sector for momenta much smaller than the nuclear scale set by
typical binding momentum in nuclei of about $100~$MeV$/c$.
The latter serves itself as basis for cluster theories which introduce composite nuclei,
like $\alpha$ particles or deuterons, for the description of larger systems.
This understanding is at present incomplete as a
reconciliation of differing power-counting schemes which define \ce~is yet to be found.
A fully renormalization-group invariant version of \ce, for example, has not yet been formulated.
However, given a reliable counting scheme and thereby uncertainty estimates will enable us to relate
the five relevant QCD and QED parameters (three light-quark masses and the
electromagnetic and strong coupling constants) via the matching of QCD and EFT
amplitudes to observables, which
are poorly known experimentally and would thus be an ultimate assessment of whether
or not \textit{the} nuclear theory has been found.

The crucial calculation of multi-hadron amplitudes with QCD became reality, recently,
via the numerical solution of the QCD path integral in discrete space-time (lattice QCD or LQCD):
unquenched, high-statistics LQCD measurements are available for the nuclear spectrum up to $A=4$ using
\textit{(i)} $N_f=2+1$, \ie, three-flavor QCD with the physical strange-quark mass and equal up,
down-quark masses $m_u=m_d$ corresponding to $m_\pi\sim300$~MeV in Ref.~\cite{Yamazaki:2015asa}~,
450~MeV in Ref.~\cite{Orginos:2015aya}~($A\leq 2$), and 510~MeV in Ref.~\cite{Yamazaki:2012hi}~,
and \mbox{\textit{(ii)} $N_f=3$,} \ie, exact SU(3)-flavor QCD because of degenerate light-quark masses
yielding \mpill~in Ref.~\cite{Beane:2012vq}~.
While measuring at unphysically high pion masses increases the signal-to-noise ratio in lattice simulations
and allows for meaningful results, it inhibits a matching of those amplitudes to a chiral effective theory which
relies on pion masses $\lesssim 500~$MeV.
Before LQCD can probe the nuclear spectrum in this pion-mass range, its amplitudes \textit{can}
be matched\scite{Barnea:2013uqa}in the region where its area of applicability overlaps with the pionless theory.

The current status of this program, to relate QCD parameters through LQCD, \ce,
and contact theories to few-nucleon observables,
is reviewed in this article. A graphical outline is given in Fig.~\ref{fig.theoland}. It shows
areas of applicability of the various theories
in the physical parameter space spanned by the pion mass, a momentum scale $Q$, and the nucleon number $A$.
In this figure, QCD is shown to probe the entire space and
is expected to coincide with experiment on the physical (red) sheet. Without QCD solutions for nuclear observables (dashed enclosed, transparent sheets)
at physical \mpi, we begin the discussion at the intersection between experiment and \ce~(yellow sheet) in Sec.~\ref{sec.lqcd2chi}, before showing how the overlap
between \ce~and \eftnopi~was utilized to reach larger nuclei (green squares indicate those where \eftnopi~has been used) in Sec.~\ref{sec.chi2nopi}~.
Before elaborating on the interface between LQCD and two contact theories (\eftnopi~amongst them) in Sec.~\ref{sec.hal} and~\ref{sec.left}, we insert with Sec.~\ref{sec.qcddata}
the status of the exploration of the two sheets at 510 and 806~MeV via the lattice method. The lattice technology is briefly touched in Sec.~\ref{sec.lqcdmeth}~.
We include comments on the further investigation of the $A$ axis by matching contact to cluster theories (cyan)
in the outlook.

\section{Lattice QCD $\to$ \ce}\label{sec.lqcd2chi}
To predict LQCD observables at physical \mpi, a theory is needed which is consistent with QCD at physical
\textit{and} some larger \mpi, where it can be matched to available LQCD data. Furthermore, the power counting
of that theory has to be understood for all values of the pion mass between the matching point and the
physical point. The extension of chiral perturbation theory to multi-baryon systems, hereafter referred to as
\ce~(concepts in Refs.~\cite{WEINBERG19913,Bedaque:2002mn}~, and
reviews in Refs.~\cite{Machleidt:2011zz,Epelbaum:2012vx})~is the only ansatz
which has been used for such an extrapolation.
The power counting of such an extension\footnote{Ref.~\cite{Fleming:1999ee}~demonstrates the complexity
of finding inconsistencies in a perturbative treatment
of the pion exchange; Ref.~\cite{Beane:2001bc}~reviews also
problems of other counting schemes.},
however, is still under development
(consider Refs.~\cite{Nogga:2005hy,Valderrama:2009ei,Valderrama:2011mv,Long:2011xw,Long:2012ve}~
next to a solution suggested in Refs.~\cite{Epelbaum:2012ua,Epelbaum:2015sha}).
Given the development of such an effective field theory for few-nucleon systems
and the availability of LQCD data at a \mpi~where chiral perturbation theory (ChPT)
converges\footnote{Ref.~\cite{Durr:2014oba}~hints to a breakdown at
$m_\pi<500~$MeV of the chiral expansion for $f_\pi$.},
the algorithm (original formulation in Ref.~\cite{Beane:2001bc}~) reviewed below can be
used to predict physical observables from LQCD data at unphysical $m_\pi$.

The chiral Lagrangian depends explicitly and implicitly on the pion mass. The implicit
dependence resides in the pion decay constant $f_\pi$,
the axial coupling constant $g_A$, and the nucleon mass $\mn$.
Using ChPT, this dependency can be made explicit in a power expansion, \eg, for the decay
constant\scite{Langacker:1973nf}with a renormalization scale $\mu$:
\begin{equation}\label{eq.pidec}
f_\pi(m_\pi)=f_\pi(m_\pi=0)\left(1+\frac{m_\pi^2}{8\pi^2f^2_\pi(m_\pi=0)}(-\ln\frac{m_\pi^2}{\mu^2}+\mathcal{O}(1))+\mathcal{O}(m_\pi^4)\right)\;\;.
\end{equation}
The \ce~Lagrangian contains in addition to single-nucleon, single-meson, and meson-nucleon coupling
parameters pion-mass \textit{independent} and \textit{dependent} contact operators\footnote{with $\mathcal{M}\sim\xi^\dagger\,\text{diag}(m_u,m_d)\,\xi^\dagger+\xi\,\text{diag}(m_u,m_d)\,\xi$, $\xi=1+\frac{i}{2f_\pi}\bs{\tau}\cdot\bs{\pi}-\frac{1}{8f_\pi^2}\bs{\pi}^2+\mathcal{O}(\bs{\pi}^3)$, Pauli isospin matrices $\bs{\tau}$, pion isovector $\bs{\pi}$, and nucleon iso-doublet $N={p\choose n}$},
\begin{eqnarray}\label{eq.chptlgr}
\mathcal{L}_\text{\small four-fermi}&=&\;\;\;C_S\left(N^T N\right)^2+C_T\left(N^T\sigma_iN\right)^2\nonumber\\&&
+D_{S1}\left(N^T\,\mathcal{M}\,N\right)^\dagger\left(N^T N\right)+D_{T1}\left(N^T\mathcal{M}\sigma_iN\right)^\dagger\left(N^T\sigma_iN\right)\\&&+D_{S2}\left(N^T N\right)^\dagger\left(N^T N\right)\text{Tr}\,\mathcal{M}+D_{T2}\left(N^T\sigma_iN\right)^\dagger\left(N^T\sigma_iN\right)\text{Tr}\,\mathcal{M}\;\;.\nonumber
\end{eqnarray}
The associated low-energy constants (LECs) $C_{S,T}$ and $D_i$ are pion-mass independent\footnote{The pions are dynamical in \ce. In \eftnopi, in contrast, the contact LECs do depend on $m_\pi$.}.
They can be determined at any \mpi~within the convergence radius of ChPT.
From the available data at physical \mpi, $D_i$ cannot be determined independently
from $C_i$. One way to resolve this ambiguity are additional scattering experiments
of pions on deuterons or heavier targets.
Another utilizes two-nucleon observables at different \mpi~as the necessary constrains to separate the explicit
\mpi~dependence from the LECs as shown in
Eq.~(\ref{eq.chptlgr})\footnote{The method of resonance saturation as developed in Ref.~\cite{Epelbaum:2001fm} and applied, \eg, to the quark-mass sensitivity of a collection of nuclear observables relevant for big-bang
nuclear synthesis\scite{Berengut:2013nh}, approaches the problem by relating \ce~contact LECs to phenomenological boson-exchange models. As it is unknown how the uncertainties of that method can be quantified, we abstain from further elaboration.}.
One LQCD data set for a pion mass which might be in the range of applicability of \ce~was calculated
in Ref.~\cite{Orginos:2015aya}~at $m_\pi=450~$MeV. 

\begin{figure}[tb]
  \centering
  \begin{minipage}[b]{0.49\textwidth}
\includegraphics[width=1.0 \textwidth]{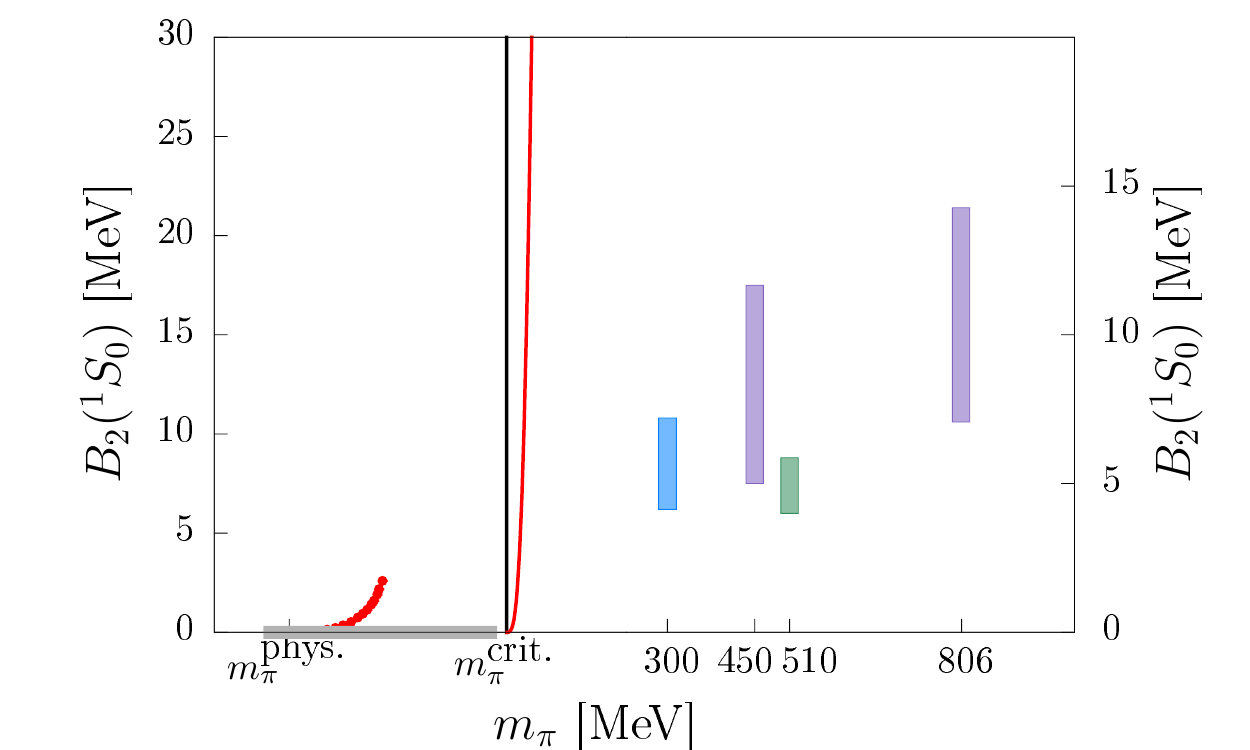}
  \end{minipage}
  \hfill
  \begin{minipage}[b]{0.49\textwidth}
\includegraphics[width=1.0 \textwidth]{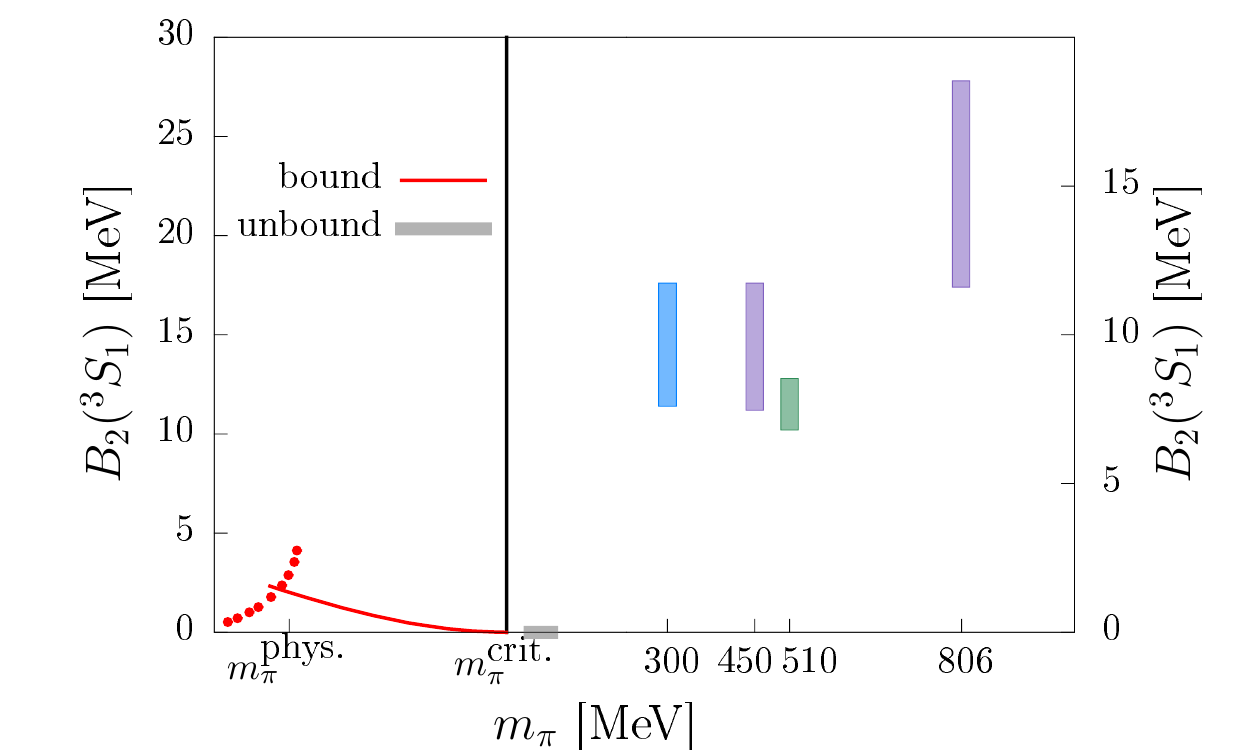}
  \end{minipage}
  \begin{minipage}[b]{0.49\textwidth}
\includegraphics[width=1.0 \textwidth]{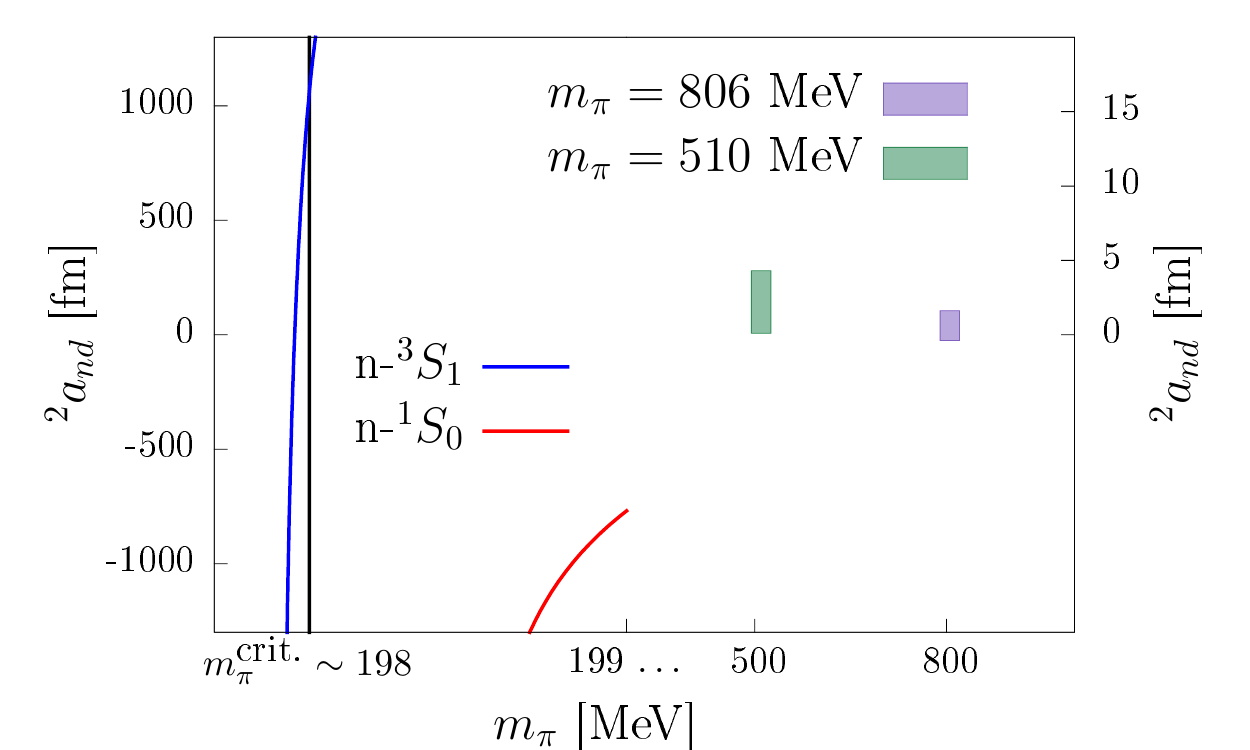}
  \end{minipage}
  \hfill
  \begin{minipage}[b]{0.49\textwidth}
\includegraphics[width=1.0 \textwidth]{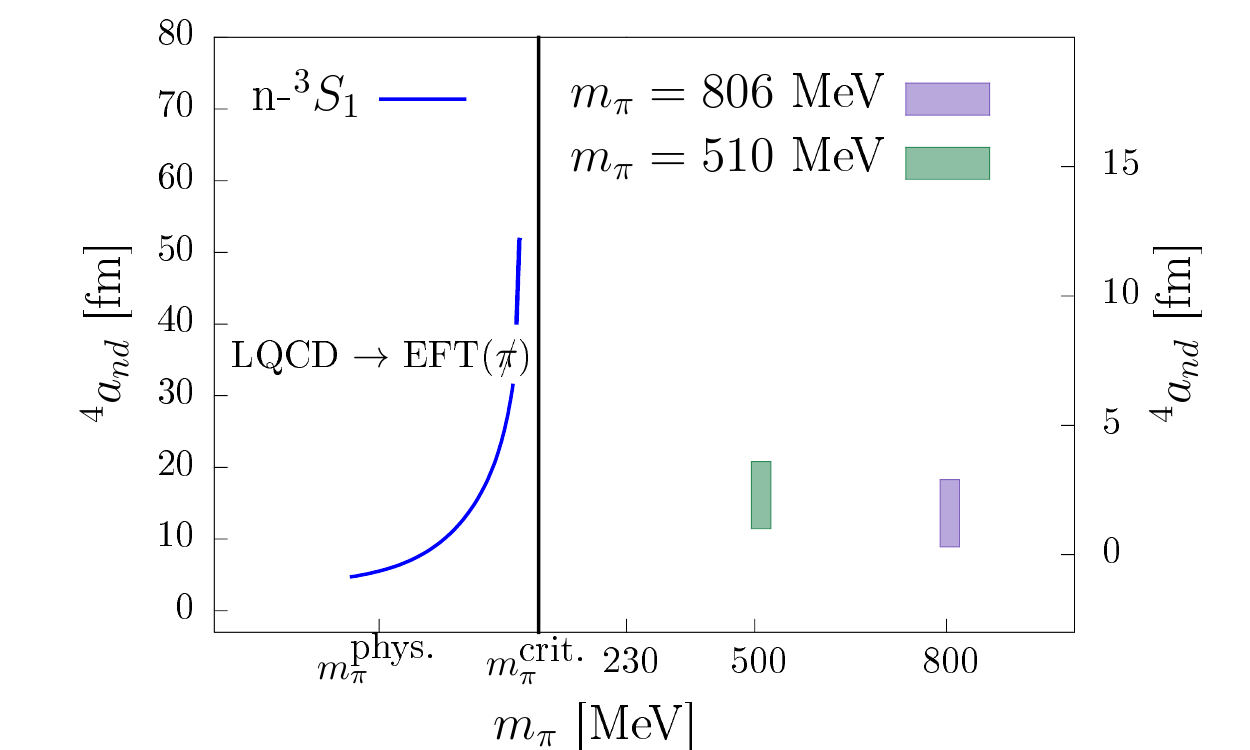}
  \end{minipage}
\caption{\label{fig.land} (Color online) Pion-mass dependence of the nn-singlet (top left)
and deuteron (top right) binding energy, the neutron-deuteron quartet-channel
scattering length (bottom right), and the neutron-deuteron(nn-singlet) (blue (red), bottom left) 
doublet scattering length. Results for \mbox{\mpi$<220$~MeV} (left y scale) represent predictions with
\eftnopi~matched to
\ce~(bound states are red solid lines\scite{Epelbaum:2006jc}and red circles\scite{Chen:2010yt},
unbound intervals are gray lines\scite{Epelbaum:2006jc})~and
for \mpi$>230$~MeV (right y scale) lattice data for $B_2$ and predictions of \eftnopi~matched to LQCD data
(see sections~\ref{sec.qcddata} and~\ref{sec.left}). Box heights correspond to lattice uncertainty and
the vertical black line marks $m_\pi\sim 198~$MeV at which $\anpst\to\infty$. All x scales are nonlinear.}
\end{figure}

Replacing LQCD data at another \mpi~with experimental data to pin down the LECs,
was demonstrated in Ref.~\cite{Chen:2010yt}~.
The authors employed the Beane-Kaplan-Vuorinen (BKV) power-counting scheme\scite{Beane:2008bt},
a variant of the Kaplan-Savage-Wise (KSW) scheme\scite{Kaplan:1998tg}
for \ce~to relate the two-nucleon amplitude to the pion mass.
The BVK scheme introduces in contrast to KSW an additional fictitious heavy meson of mass $\lambda$.
In effect, this particle cancels the $r^{-3}$ singularity of the pion exchange in the triplet
channel where KSW does not converge. The mechanism is similar to the Pauli-Villars regulation
in QED.
The scattering lengths at physical $m_\pi$ and at $\sim350~$MeV of Ref.~\cite{Beane:2006mx}~were used to 
renormalize the NN
counter terms unambiguously. The pion-mass dependence of the effective range can be inferred from data
at a single $m_\pi$ because up to NLO, operators
proportional to the quark mass do not contribute\scite{Chen:2010yt}. In addition to the ensuing $m_\pi$
dependence of the deuteron binding energy (as a pole of the amplitude Eq.~(\ref{eq.pole})),
Ref.~\cite{Chen:2010yt}~derived the charge radius, the magnetic moment and polarizability,
and the photodisintegration of the deuteron as a function of the pion mass. We compare the binding
energies in the singlet and triplet channels of Ref.~\cite{Chen:2010yt}~in Fig.\ref{fig.land} 
(thin red line) with the results obtained with a chiral potential\scite{Epelbaum:2006jc}.
The latter neither binds the singlet state at a higher $m_\pi$, nor does it indicate $\lim_{m_\pi\to 0}\eodim=0$.

Before Ref.~\cite{Chen:2010yt}~, and without lattice data at sufficiently low pion masses,
\ce~was used, also with a modified KSW counting which uses assumptions about details of
the short-range structure of the interaction, for an extrapolation in \mpi~in Ref.~\cite{Beane:2001bc}.
Like BVK, this Beane-Bedaque-Savage-van-Kolck (BBSvK)
scheme is identical to KSW in the NN singlet channel but
as an expansion about the chiral limit $m_\pi=0$ iterates the pion exchange in the coupled triplet
channel (Weinberg counting).
In Ref.~\cite{Beane:2001bc}~, BBSvK is used to investigate the deuteron and ${}^1S_0\;,\;{}^3S_1$
neutron-proton ($np$) scattering properties for pion masses from 0 up to $200~$MeV.
The two-nucleon amplitude was matched at physical \mpi~to
the triplet $np$~scattering length ($\anpt$) and effective range ($r_3$)
to determine the relevant contact terms in Eq.~(\ref{eq.chptlgr}).
With a leading order chiral expansion for $f_\pi$, $g_A$, and $\mn$,
the deuteron was found to become unbound\footnote{The value was also found\scite{Beane:2001bc} to be
quite sensitive to the regularization of the 
contact interactions.} at $m_\pi\sim 100~$MeV with a corresponding divergent $\anpt$.

The extension of the method to the $np$-singlet channel\scite{Beane:2002vs}
demonstrated the relevance of the above
ambiguity in the contact terms. One choice of $C_i$ and $D_i$ left
the di-neutron ($nn$) unbound for all pion masses in
$[0,300~\text{MeV}]$ while another yielded $nn$ bound in the interval between $180$ and $240~$MeV.
The problem was then addressed in Ref.~\cite{Beane:2002xf} by placing
constrains on the contact terms with na\"ive dimensional analysis.
A relevant conclusion for the extrapolation from $m_\pi>140$~MeV down
was the consistency of a bound $nn$, \ie, singlet two-nucleon
state, with data at the physical pion mass and coupling strengths of natural size.

The effect a different power counting has on the \mpi~dependence of the two-nucleon system is seen when comparing
the results of Refs.~\cite{Beane:2001bc,Beane:2002vs,Beane:2002xf}~, to Ref.~\cite{Epelbaum:2002gb}~, and to
Ref.~\cite{Soto:2011tb}~.
Considering also two-pion-exchanges in a modified\scite{Epelbaum:1998ka}Weinberg counting, Ref.~\cite{Epelbaum:2002gb}~found the deuteron in the chiral limit more strongly bound than in nature.
While this is in contrast to the unbound deuteron of Ref.~\cite{Beane:2001bc}~,
the absence of bound states in the singlet $NN$
channel for $m_\pi\in(0,200~\text{MeV})$ is a common result of both schemes.
pions in Ref.~\cite{Soto:2011tb}~finds the singlet and deuteron state unbound in the chiral limit.
This work employs composite two-nucleon spin-singlet and triplet quasi-particles
(dibaryons, see Sec.~\ref{sec.left}) coupled to single nucleons and propagates
the pion-mass dependence to a pionless theory through the analytical matching of amplitudes.
These differences should concern convergence rates and precision of the EFT but must eventually lead
to the same observables. This issue is unresolved.
As a feature, ref.~\cite{Soto:2011tb}~demonstrates a reversed extrapolation from the physical
pion mass to larger values. The LQCD results which are available (see Sec.~\ref{sec.qcddata}) there
provide support that nature as represented by experimental data at physical $m_\pi$ ``flows''
into SU(3)-large-$m_q$ QCD as explored via the lattice method.

Regardless of the counting scheme, a critical pion mass close to the physical value was found where the
two-body scattering lengths diverge. The associated scale invariance of the two-nucleon system \wrt~
any coordinate transformation $\bs{r}\to\lambda\,\bs{r}$ for $\lambda\in\mathbb{R}_+$ indicates
an infrared fixed point of the renormalization-group (RG)
flow of the QCD coupling constant\footnote{This nontrivial inference assumes a valid approximation of QCD by
\ce. Only then does a statement based on a \ce~calculation about a QCD parameter make sense.}
$g_s$.
How this \textit{infrared} QCD fixed point is expressed in the RG flow of \ce~is
unknown and presents a problem intimately related to the development of a consistent power counting.
In contrast, the flow is known for an effective theory that considers solely contact interactions amongst nuclei
(see Sec.~\ref{sec.chi2nopi} below). Although not manifest in its coupling constants, the peculiar limit-cycle
trajectory of the leading three-body LEC, emergent from the fixed point,
should find its signature in the \ce~spectrum.
Part of this signature, namely remnants of an Efimov spectrum\scite{Efimov:1971zz},
was found\scite{Epelbaum:2006jc}when the authors employed an interaction derived
in Ref.~\cite{Epelbaum:2002gb}~in the three-nucleon system.
The characteristic asymptotic ratio of binding energies at the deuteron-neutron accumulation point of $\sim 515$
was approximated, while the ground state, \ie, the triton was stable ($\eotri=4\pm2~$MeV)
in the considered pion-mass interval between 190 and 210~MeV.
The \ce~predictions in Fig.~\ref{fig.spectrum} were obtained at a fixed cutoff
and thus do not allow an assessment of the uncertainty
like the displayed LQCD and physical experiments.

In essence, all three approaches\scite{Beane:2001bc,Epelbaum:2002gb,Soto:2011tb}employ \ce~to define
a Lagrangian with implicit and explicit dependence on \mpi. They differ in the power counting
and thereby the dependence of observables on the parameters of $\mathcal{L}$, \ie, \mpi.
Pion-deuteron scattering experiments in the physical or lattice world separating the $D_i$ counter terms from the $C_i$'s,
like $(N^T N)^2\pi\pi$, or LQCD $(N^T N)^2$ amplitudes at different \mpi~will allow for a unique determination of the LECs,
and thereby aid the development of the above mentioned consistent power counting.

To that end, the EFT analysis in Ref.~\cite{Orginos:2015aya}~demonstrated a match
of \ce~to LQCD instead of experimental data similar to Ref.~\cite{Chen:2010yt}~.
In the KSW\scite{Kaplan:1998tg}and
BBSvK\scite{Beane:2001bc}counting schemes, this work considered the EFT convergence in the
${}^1S_0$ and ${}^3S_1$ scattering lengths and effective ranges up to NNLO with LQCD binding energies
and phase shifts at $m_\pi=450~$MeV as input. The convergence rate was found small in all 4 observables
and different from the ones extracted from the LQCD phase shifts via an effective range formula. The latter
discrepancy was explained\scite{Orginos:2015aya}by fitting to data beyond the natural scale of the theory. 

The results at 450~MeV pion mass are entering a region where they can be analyzed with low-energy theorems
as developed in Ref.~\cite{Baru:2015ira}~. These theorems provide expansions of the modified
effective range parameters in $m_\pi/m_\rho$. This assumes a long-range interaction provided by \ce, hence \mpi~
as the light scale, and a short-range potential of range $m_\rho^{-1}$.
The uncertainty due to an unknown $m_\pi$ dependence on the short-range part of the interaction
was found\scite{Baru:2015ira}to increase significantly with $m_\pi$.
%
%
Part of this uncertainty is, again, the isolation of
the explicit pion-mass dependence in the four-fermi contact
LECs\footnote{The increasing effect with \mpi~of the unknown
$m_\pi$ dependence of the short-range interaction is visualized in Ref.~\cite{Baru:2015ira}
which in turn will also serve as a measure of future rigorous $C_i$ and $D_i$ assignments.}.
The method applied to $m_\pi\leq 400~$MeV
yields results for $a/r$ which are consistent within error bars
with the \mpif~data (see Table~\ref{tab.latdata_1} and
compare Fig.~\ref{fig.aland} with Figs.~6 and 7 in Ref.~\cite{Baru:2015ira}~) regardless of being
extrapolated from the physical \mpi~or constrained by LQCD deuteron data directly at
$m_\pi=430~$MeV. Even a na\"ive continuation of the derived \mpi~dependence of the ratio $a/r$ is
``in line'' with the peculiar value of 2 LQCD finds at 806~MeV pion
mass (see discussion around Eq.~(\ref{eq.pole})).

The above attempts employ effective field theories to parameterize two- and three-nucleon amplitudes, \ie,
the scattering lengths in singlet and triplet S-wave channels $\anpst$ and the triton binding energy $\eotri$,
with the pion mass. The pion mass and the theoretical uncertainty of \ce~can then be passed
on to another effective theory which is more practical for larger nuclear systems.
This matching between a pionful and pionless theory cited above in Refs.~\cite{Chen:2010yt,Soto:2011tb}~,
leads to the
next section where we summarize work which already employed the idea to propagate the pion-mass dependence
from two- into few-nucleon amplitudes with the caveat of an unknown uncertainty in those amplitudes and therefore
\mpi~dependence of the underlying theory which is then no longer QCD but \ce.

\section{\ce~$\to$ \eftnopi}\label{sec.chi2nopi}
Above, we reviewed the derivation of the low-energy effective theory \ce~from its underlying theory
QCD for a practical description of few-baryon systems at a momentum scale $\sim\mathcal{O}(m_\pi)$.
In this section, we summarize results of a contact effective theory without pions (\eftnopi, see Sec.~\ref{sec.left} for references).
Its underlying fundamental theory is \ce~from which it inherits the QCD-parameter dependence.
Because of the stated unresolved problems of \ce~to quantify its theoretical uncertainty
those errors are also not properly propagated to \eftnopi.
The set of observables, thereby parameterized by couplings of
QCD\footnote{The one-dimensional pion-mass dependence will be augmented once electromagnetism
and non-degenerate up and down-quark masses are considered in the lattice simulations.}
\textit{and} the matching conditions, is smaller than that accessible with \ce~ which in turn does not cover the measurable
parameter space described by QCD.
Matching conditions, in general, encompass:
\mbox{\textit{(i)} The point} in the QCD parameter space where amplitudes are equated, \ie, the relevant QCD interaction parameters
($m_q$, strong and electromagnetic couplings $g_s$ and $\alpha$);
\textit{(ii)} A set of consistent quantum numbers specifying
the observables which are expected to be well described in either theory
(\eg, the deuteron, $\anpt$, and/or the magnetic moment of the triton);
\textit{(iii)} A parametrization of the radius of convergence of the effective
theory (power counting).
\begin{figure}[tb]
\centerline{\includegraphics[width=1. \textwidth]{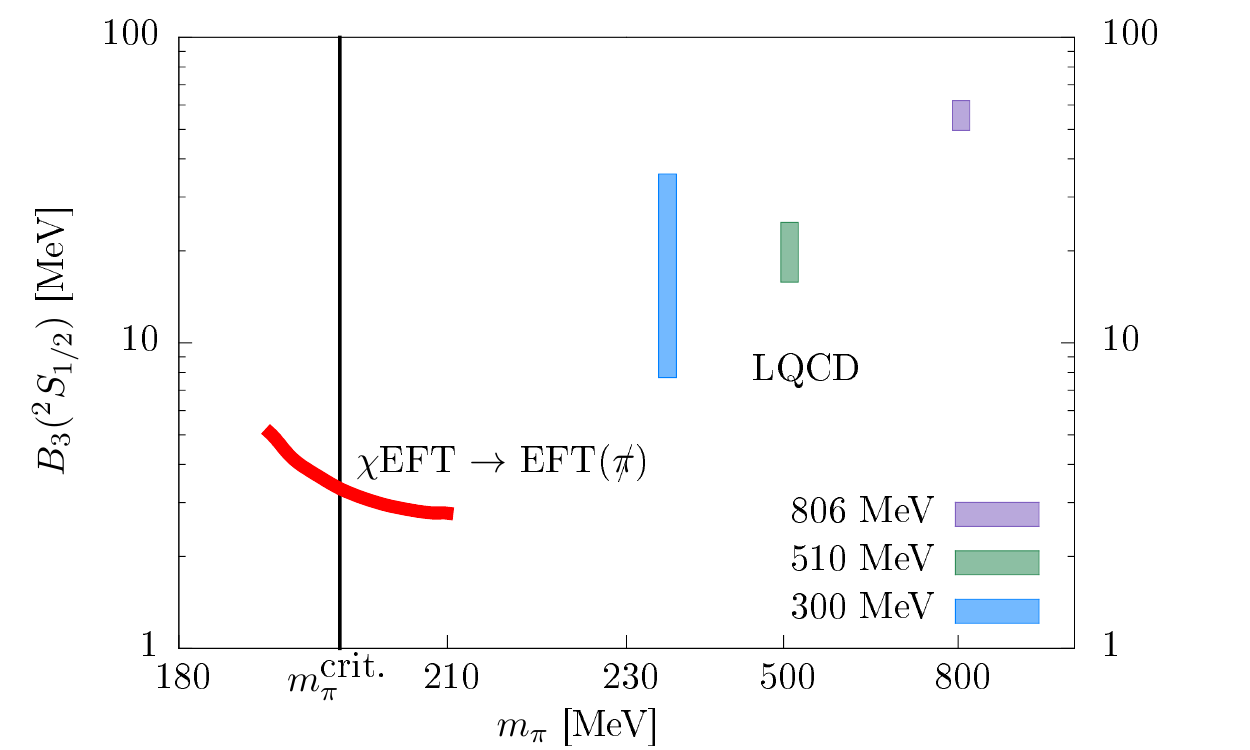}}
\caption{\label{fig.3bdyland} (Color online) Pion-mass dependence of the three-nucleon binding energy in the triton channel.
For \mpi$<210~$MeV, the results\scite{Epelbaum:2006jc}utilize a match of \ce~to data at physical \mpi. 
For \mpi$\gtrsim 300~$MeV, LQCD data is shown with box heights representing the lattice uncertainty 
(see Sec.~\ref{sec.qcddata}). The vertical black line marks $m_\pi\sim 198~$MeV at which $\anpst\to\infty$.}
\end{figure}

The resultant benefit of a practical theory to express few-nucleon observables in terms
of QCD parameters was utilized\scite{Epelbaum:2006jc}for the prediction of the three-nucleon
spectrum for $m_\pi\in(140,200)~$MeV.
In that work, the match to \eftnopi~was made to predict three-nucleon scattering
observables and excited states of the triton. Prior to Ref.~\cite{Epelbaum:2006jc}~,
the ultraviolet RG fixed point in the two-body and the limit cycle in
the three-body sector of \eftnopi~was conjectured at a pion mass close
to the physical point\scite{Braaten:2003eu}
under the assumption that an amplitude relevant for the description of the three-nucleon
system\footnote{Originally coined~$\Lambda^*$ in Ref.~\cite{Bedaque:1998kg}~.}
remains constant over the considered range of pion
masses which includes the physical point from where its value was determined.
This assumption was justified \textit{a posteriori} in
Ref.~\cite{Epelbaum:2006jc}~by solving the three-nucleon problem with a \ce~potential and using the obtained
pion-mass dependence of $\eotri$ to obtain the aforementioned first two states of an Efimov spectrum.
The limit cycle as a periodic dependence of \textit{a} coupling constant\footnote{Only in
the doublet channel, this constant can be identified with
a momentum-independent six-fermi vertex. In the quartet channel, any six-fermi
counter term has to be momentum dependent or break (iso)spin symmetry to affect the quartet S-wave amplitude.}
is manifest in the neutron-deuteron
scattering lengths. In Ref.~\cite{Epelbaum:2006jc}~the neutron-deuteron scattering length was
found to diverge in \textit{both} spin channels at a critical pion mass of $\sim198~$MeV
with \eftnopi~at LO. This is a noteworthy difference to the nucleon-deuteron system at physical \mpi, where
\eftnopi~predicts a limit cycle only for the doublet channel and the three-body momentum-dependent counter
terms of the quartet channel appear at an order expected by dimensional analysis.
An ensuing NNLO analysis\scite{Hammer:2007kq}confirmed this behavior of
the scattering lengths showing traces of a log-periodic behavior in the vicinity of
the critical pion mass of about 198~MeV. The dependence of the nuclear landscape
on $m_\pi$ ranging from 130~MeV to 806~MeV
up to $A=3$, the two complementary studies of Refs.~\cite{Epelbaum:2006jc,Hammer:2007kq}~
give in combination to the analyses presented below in Sec.~\ref{sec.left},
is shown in Figs.~\ref{fig.land} and~\ref{fig.3bdyland}~.

In the upper row of Fig.~\ref{fig.land}, the singlet (left) and triplet (right) $NN$ binding energies are shown.
The \ce~calculations extrapolate from the physical \mpi~up to $\sim 210~$MeV and find
a critical pion-mass value where scattering lengths in both channels diverge\footnote{The
LECs were deliberately chosen to
that end. Singlet and triplet scattering lengths could diverge at different \mpi.
That magnetic fields could be tuned to realize diverging scattering lengths in all
two-nucleon channel was suggested in Ref.~\cite{Detmold:2015daa}.}
at $m_\pi\sim 198~$MeV. At this value, the deuteron becomes unbound and
the singlet-$nn$ bound. The LQCD calculations
(for references and numerical values see Table~\ref{tab.latdata_1}) at
300, 510, and 806~MeV find more deeply bound states
in both channels. This peculiar situation requires at least one more \mpi~at which $\anpt$ diverges but
allows multiple scenarios for the singlet: \textit{(i)} no more critical point and a gradual increase from the
shallow state (red thick line upper left) to the deeper states; \textit{(ii)} a intrusion of another bound state at
threshold at some critical \mpi~where the former ground state is deeply bound. Both cases indicate the
existence of one or more pion masses $210~\text{MeV}<m_\pi<300~$MeV where QCD is on the RG trajectory for
two-nucleon ultraviolet fixed points and hence a three-nucleon limit cycle. To observe signatures of this critical
trajectory in LQCD simulations could thus be feasible, today.

What is known about the \mpi~dependence of the triton binding energy
directly (LQCD) and indirectly ($\chi$EFT) from QCD is
shown in Fig.~\ref{fig.3bdyland}. Around the critical point, Ref.~\cite{Epelbaum:2006jc}~finds the ground state
stable (red line) relative to the increase observed in LQCD data from 300 to 806~MeV. The above-mentioned tower
of excited states which forms around $m_\pi^\text{crit.}$ is not shown. Instead, we show the manifestations
of the unitary fixed point in the three-body sector,
as calculated in Refs.~\cite{Epelbaum:2006jc,Hammer:2007kq}~
(\ce$\to$\eftnopi) and Ref.~\cite{Kirscher:2015yda}~(LQCD$\to$\eftnopi), in the bottom row of Fig.~\ref{fig.land}.
For $m_\pi<198~$MeV, the doublet scattering length $\andd$~(left) is obtained from neutron-deuteron asymptotic states.
$\andd$~diverges at the critical point to $+\infty$. Above the critical pion mass, the deuteron is unbound and
is replaced with the then bound nn-singlet state in the definition of $\andd$. The latter rises from $-\infty$.
In the neutron-deuteron quartet channel (right lower panel), the scattering length $\andq$ goes to $+\infty$ at
the critical point and cannot be defined in the gap between $m_\pi^\text{crit.}$ and some \mpi~below 300~MeV
where the $^3S_1$ state is bound, again.
At present, we know of no systematic way to connect the \ce~predictions with the LQCD measurements. To understand
the emergence and disappearance of bound states with \mpi, LQCD calculations tracing, \eg, the two-nucleon spectrum
over some range of \mpi, would indicate whether the shallower state at 510~MeV corresponds to the lowered pole
at 806~MeV or enters at threshold. The statistical uncertainty is the main obstacle to answer the
seemingly simple question how $\eodim$ reacts to an infinitesimal change in \mpi.
The slope of $\eodim$ as a function of \mpi~could be obtained for $m_\pi\geq~806~$MeV
and provide valuable information for the extrapolation to smaller pion mass.

We will see below (discussion around Fig.~\ref{fig.lecs}) hints
that the nuclear interaction at large \mpi~is well 
approximated by a SU(4) Wigner-symmetric theory.
Assuming that \ce~propagates the explicitly broken SU(2) or SU(3) quark-flavor symmetry well into the interaction
between nuclei, the \ce~results are not consistent with the LQCD results,
which use degenerate up and down quark masses.
This discrepancy highlights the significance of the symmetry breaking for an extrapolation.

In this section, we summarized results of \ce~and \eftnopi~as approximations of QCD. 
Bound and scattering properties of nucleon systems with $A\leq3$ were found sensitive to the QCD-parameter \mpi.
This behavior mediated by \ce~was obtained
without a systematic power counting and ambiguities due to insufficient data.
How a contact theory can complement \ce~in the facilitation of scattering calculations while
converging in its predictions of the three-nucleon spectrum was also shown.
Until now, \ce~inherits only its symmetries from QCD. The LECs are determined
by a match to experimental data at $m_\pi\sim 140~$MeV because LQCD cannot be solved for the necessary amplitudes
for $m_\pi\lesssim 300~$MeV. 
However, the variety of nuclear behavior to expect when substituting the scattering lengths and effective ranges
at physical \mpi~with LQCD predictions at some $m_\pi\lesssim 200~$MeV where ChPT seems to be applicable as matching
conditions, was exemplified.

The analysis of nuclei at larger \mpi, where ChPT is not a reasonable approximation of QCD but where LQCD is
practical, today, is the subject of the following sections.
\section{Lattice QCD for multi-hadron systems}\label{sec.lqcdmeth}
To predict nuclear observables rigorously by solving QCD, a chain of systematic approximations has to be employed. Each component of this chain is built as an effective field theory and as such provides a prescription to recover the
underlying theory. The belief in QCD as the relevant theory from which nuclei emerge can be tested by 
calculating observables from correlation functions of the type
\begin{equation}\label{eq.pathint}
\langle\hat{O}\rangle\equiv\mathcal{Z}^{-1}\int\mathcal{D}A_\mu\mathcal{D}q\mathcal{D}\overline{q}\;\hat{O}(q,\overline{q},A_\mu)\,e^{-\int d^4x\mathcal{L}_\text{QCD}(m_q,g_s)}\;\;.
\end{equation}
The lattice methodology evaluates this path integral and the partition function $\mathcal{Z}=\langle 1\rangle$
over fermionic quark fields ($q,\overline{q}=q^\dagger\gamma^0$) and gluon gauge fields ($A_\mu$) by discretizing Euclidean space time. Thereby, LQCD constitutes the
first approximation~---~or chain element in the above terminology~---~to relate QCD parameters, namely the quark masses $m_q$ and the QCD length scale, implicit in the coupling strength $g_s$, to nuclear physics.

For few hadron systems, in fact, it is the only approximation necessary. For spectral details beyond the ground state, larger systems, where $\hat{O}$ is comprised of \mbox{$\geq 5$} baryons, and/or realistic values
of $m_q$, additional EFTs are necessary. The various EFTs, ChPT, and \eftnopi, for instance, interface through amplitudes with $\hat{O}$ resembling an operator that can be evaluated in both, the underlying and effective theory.
\subsection{Methodology}\label{sec.latmet}
First, the defining parameters of a lattice calculation, roughly taken as the spacing $b$ of the lattice, its
space-time volume $L^3\times T$, the statistical properties of the gauge-field ensemble, and the $m_q$, $g_s$
values affect the accuracy of the extraction.
Furthermore, an appropriately chosen shape for the operator $\hat{O}$ increases the
accuracy.
To access spectra of few-nucleon systems,
$\hat{O}$ is composed of hadronic interpolating fields with the generic structure
\begin{equation}\label{eq.hadrint}
\overline{N}^h=\sum_{\bs{a}}\Gamma_h^{a_1,\ldots,a_{n_q}}\overline{q}(a_1)\ldots\overline{q}(a_{n_q})\;\;\;,\;\;\;a_i=\lbrace\textrm{color, flavor, spinor, spatial}~\bs{x}\rbrace\;\;.
\end{equation}
A hadron is identified with a certain irreducible representation of SU(3) flavor, \eg, $n,p,\Lambda,\Sigma^{\pm,0},\Xi^\pm\leftrightarrow\mathbf{8}$, and therefore neither the number of quark fields $n_q$ nor the structure of the tensor 
$\Gamma_h$ are uniquely defined by the hadron. Spectroscopic information on hadrons is obtained through the selection of an appropriate operator $\hat{O}$, which creates a state of defined parity $\pi$, cubic angular momentum $l$, isospin,
strangeness at some initial space-time point and annihilates it later. The so called smearing of the quark fields to increase the overlap of the interpolating fields with an eigenstate of the QCD Hamiltonian via
\begin{equation}\label{eq.smear}
q(\bs{x},t)=\sum_{\bs{y}}G(\bs{x},\bs{y},A_\mu(t))q(\bs{y},t)
\end{equation}
utilizes this freedom (the calculations considered in Table~\ref{tab.lqcdpara_1} differ in the weight function $G$ and to which parts of $\hat{O}$ the technique is applied).

All lattice calculations under review, here, employ a \textit{source} and \textit{sink} structure for $\hat{O}$. States are created with parity $\pi$, total angular momentum $J^2$ and $J_z$,
isospin $I^2$ and $I_z$, strangeness, and baryon number $A$ at the \textit{same}
time $t_0$ to annihilate them at a $t>t_0$. To access the spectrum of a three-baryon system like the triton or Helium-3, for example, correlation functions of the form 
\begin{eqnarray}\label{eq.3bcorr}
{}\hfill C_{N_1N_2N_3;\Gamma}(\bs{p}_2,\bs{p}_1,\bs{p}_3;t)=\sum\limits_{\bs{x}_1,\bs{x}_2,\bs{x}_3}e^{i\bs{p}_1\cdot\bs{x}_1}e^{i\bs{p}_2\cdot\bs{x}_2}e^{i\bs{p}_3\cdot\bs{x}_3}\Gamma_{3;\alpha_1\alpha_2\alpha_3}^{\beta_1\beta_2\beta_3}\times
\nonumber\\
\times\left\langle N_1^{\alpha_1}(\bs{x}_1,t)N_2^{\alpha_2}(\bs{x}_2,t)N_3^{\alpha_3}(\bs{x}_3,t)\overline{N}_{1,\beta_1}(\bs{x}_0,0)\overline{N}_{2,\beta_2}(\bs{x}_0,0)\overline{N}_{3,\beta_3}(\bs{x}_0,0)\right\rangle
\end{eqnarray}
are used\scite{Beane:2009gs}with spinor indices $\alpha,\beta$. The particle species is understood to be 
encoded in the $N_i$'s.
The space-time points at which the \textit{source} creates ($\bs{x}_0,t_0=0$) and the
\textit{sink} annihilates ($\bs{x}_i,t$) hadron(s) with
quantum numbers selected through $\Gamma$ are choices like
the projection on a state with defined momenta $\bs{p}_i$.
Another choice are so called wall sources used, \eg, in Ref.~\cite{Ishii:2006ec}~, which
project each quark field of zero momentum out of the vacuum \textit{viz.} $q(t)=\sum_{\bs{x}}q(\bs{x},t)$.

So far, nuclear observables have been calculated with fields $N$ resembling elements of the baryon octet, only.
In Ref.~\cite{Beane:2012vq}~, the proton mass is obtained with an $A=1$ interpolating field $\overline{N}^\alpha(\bs{x},t)=\epsilon_{abc}(u^{a,T}C\gamma_5d^b)u^{c,\alpha}(\bs{x},t)$ corresponding to a color (indices $abc$)
singlet which is combined from up ($u$) and down ($d$) quark fields defined at the same space-time point but smeared (Eq.~(\ref{eq.smear})), and the charge conjugation matrix $C$ acting on
the spinor components (index $\alpha$).
At the hadronic level in Eq.~(\ref{eq.3bcorr}) one chose to define the three sink and source operators at
the same space-time point. In general, smearing these operators over the spatial coordinates
is admissible, too, like shown in Eq.~(\ref{eq.smear}) for the individual quarks.

Back on the hadronic level this dependence on individual quark coordinates is necessary
to extract not only spectral information, but also wave functions.
The standard technique to access both, wave functions $\Psi(\bs{\rho}_{1,\ldots,A-1};k,t)$
and energy eigenvalues
$E_n=2\sqrt{k^2+\mn^2}$, first translates the sink
\mbox{$N_s(\bs{x},t)=e^{H_\text{QCD}t}N_s(\bs{x},0)e^{-H_\text{QCD}t}$} in time
before the insertion of a complete set of states:
\begin{eqnarray}\label{eq.efmf}
C(\lbrace\bs{\rho}_i^\text{  s(ource),si(nk)}\rbrace_{i=1,\ldots,A_1})\propto\hspace{7cm}\nonumber\\
\propto\left\langle N_\text{si}(\lbrace\bs{\rho}_i^\text{  si}\rbrace_{i=1,\ldots,A-1},t)\overline{N}_\text{s}(\lbrace\bs{\rho}_i^\text{  s}\rbrace_{i=1,\ldots,A-1},t_0=0)\right\rangle\hspace{3.2cm}\nonumber\\
\propto\sum\limits_n\frac{e^{-E_nt}}{2E_n}\left\langle N_\text{si}(\lbrace\bs{\rho}_i^\text{  si}\rbrace_{i=1,\ldots,A-1},t)\left\vert n\left\rangle\right\langle n\right\vert\overline{N}_\text{s}(\lbrace\bs{\rho}_i^\text{  s}\rbrace_{i=1,\ldots,A-1},t_0=0)\right\rangle\hspace{0.5cm}\nonumber\\
\to Z_n\left(\lbrace\bs{\rho}_i^\text{  si}\rbrace_{i=1,\ldots,A-1};k\right)Z_n\left(\lbrace\bs{\rho}_i^\text{  s}\rbrace_{i=1,\ldots,A-1};k\right)\Psi_0(\bs{\rho}_{1,\ldots,A-1};k,t)e^{-E_0t}\;\;.
\end{eqnarray}
For $t\to\infty$, the contribution from the lightest hadronic state that couples to source \textit{and} sink dominates. The overlap factors $Z_n$ depend, of course, on the sink/source structure and determine the minimal propagation time $t$ for a
practical extraction of $E_0$. The residual time dependence in the sink/source overlap with the eigenstates
indicates the statistical noise which is present through Monte-Carlo sampling in realistic calculations.
It is noteworthy, that $n=0$ in Eq.~(\ref{eq.efmf}) does not need to be the ground state for a given time interval. The state of interest, ground, excited, or continuum state, guides the structure chosen for the sink and
source. Smearing of quark fields in the nucleon interpolating field, for instance,
is inspired by the knowledge of the extended nature of the nucleon. For scattering states (see paragraph below) in the center-of-mass frame, the correlator with a sink structure which
projects hadrons with momenta of equal magnitude but opposite direction will be dominated by an excited state over some time interval, because the overlap factors $Z_n$ in Eq.~(\ref{eq.efmf}) win over the exponential decay.

As wave functions and energies are calculated in a box of finite size $L$, scattering phases $\delta$ can be inferred from the correlation function. If the energy extracted via Eq.~(\ref{eq.efmf}) corresponds to a scattering state with $E=2k^2/\mn$,
L\"uscher's formula (original work in Refs.~\cite{Luescher1986,Lüscher1991531}~,
for boosted systems see Ref.~\cite{Davoudi:2011md}~)
\begin{equation}\label{eq.luescher}
k\cot\delta=\frac{1}{\pi L}\sum_i^{\Lambda_i}\frac{1}{\bs{j}^2-\left(\frac{kL}{2\pi}\right)^2}-4\pi\Lambda_j
\end{equation}
with a sum over all integer three-vectors $\bs{j}$ with magnitude smaller than the lattice-momentum cutoff $\Lambda$, relates the energies to phase shifts.

The algorithm employed by the NPLQCD collaboration to calculate multi-baryon correlation
functions summarizes this section.

First (hadron level), an $A$-baryon operator is defined via Eq.~(\ref{eq.hadrint}) with flavors taken as elements of the baryon octet. All baryons comprising the source in Refs.~\cite{Beane:2009gs,Beane:2012vq}~are defined at a single lattice site,
\ie, the sum in Eq.~(\ref{eq.hadrint}) excludes spatial degrees of freedom, and the fields are created at the same $\bs{x}_0$.
The sink is chosen to project out a state with defined momentum for \textit{each} individual baryon, \ie, the sum in Eq.~(\ref{eq.hadrint}) runs over all lattice sites and the factor $\Gamma_h^{a_1,\ldots,a_{n_q}}$ contains a plane wave $e^{i\bs{p}\cdot\bs{x}}$
for each baryon. In principle, the ensuing momentum dependence of the overlap function Eq.~(\ref{eq.efmf}) provides a rigorous assessment of the typical baryon momenta within a nucleus. Measurements like this are desirable
for the construction of effective theories (see Sec.~\ref{sec.left}) as they
could provide their typical scales.

Second (quark level), the baryon operators are substituted with smeared\scite{Yamazaki:2012hi,Yamazaki:2015asa}(smeared and point\scite{Beane:2012vq}) interpolating operators, again defined at a single space-time point.

From correlation functions constructed in this way, properties of nuclei with $A\leq 4$ have been
calculated at the SU(3) flavor-symmetry point and for degenerate up and down quarks with the strange
at its physical mass (SU(2)).
We compare the parameters of those calculations in Table~\ref{tab.lqcdpara_1} using the
lattice spacing $b$, the temporal and spatial lattice size $T\times L^3$, the number of gauge configurations $N_\text{cfg}$, and the QCD version, \ie, SU(3) or SU(2) flavor,
as standard. The selected calculations supersede the pioneering
studies\scite{Yamazaki:2009ua,Yamazaki:2011nd,deForcrand:2009dh}which rely on the uncontrolled quenched approximation.
In contrast, the analyses under review in Sec.~\ref{sec.qcddata}~use
controlled approximations, only, in the sense that the usage of more appropriate parameters
as the ones given in Table~\ref{tab.lqcdpara_1} must yield results consistent
with the old predictions within error bars.
\begin{table}
\setlength{\tabcolsep}{4pt}
\renewcommand{\arraystretch}{1.4}
\caption{\label{tab.lqcdpara_1}{Parameters of the LQCD measurements reviewed in this work. Few-nucleon systems are calculated
on a $L^3\times T$ lattice with spacing $b$ using three or two degenerate quark masses with resultant pion (\mpi) and nucleon ($\mn$)
masses in an ensemble comprising $N_\text{cfg}$ gauge configurations.}}
\small\centering
\begin{tabular}{l|llcccl}
\hline\hline\rowcolor{white}
 QCD version & $L~$[fm] & $T~$[fm] & $b~$[fm] & \mpi~[MeV] & $\mn~$[MeV] & $N_\text{cfg}$ \\
 \hline
 SU(3)\scite{Beane:2012vq,Beane:2013br}  & $3.4\to 6.7$   & $6.7\to9.0$    & $0.15$ & $806$          & $1634$         & $\geq 1905$    \\
 SU(2)\scite{HALQCD:2012aa}              & $2.9$          & $5.8$          & $0.09$ & $701$          & $1583$         & $390$          \\
 SU(2)\scite{Yamazaki:2012hi}            & $2.9\to 5.8$   & $4.3\to5.8$    & $0.09$ & $510$          & $1320$         & $200$          \\
 SU(2)\scite{Beane:2006mx}               & $2.5$          & $4.0$          & $0.13$ & $354,493,593$  & $-$           & $490\to 660$   \\
 SU(2)\scite{Orginos:2015aya}            & $2.8\to 5.6$   & $7.5\to 11.2$  & $0.12$ & $450$          & $1226$         & $\geq 1000$    \\
 SU(2)\scite{Yamazaki:2015asa}           & $4.3\;\&\;5.8$ & $4.3\;\&\;5.8$ & $0.09$ & $300$          & $1050$         & $400\;\&\;160$ \\
 SU(3)\scite{Inoue:2011ai,Inoue:2014ipa} & $3.9$          & $3.9$          & $0.12$ & $469\to 1171$  & $1161\to 2274$ & $720\to 420$   \\
\hline\hline
\end{tabular}
\end{table}
\subsection{Data}\label{sec.qcddata}
The data selected for this review represents the most advanced LQCD extractions
(judged by the standard parameter set listed in Table~\ref{tab.lqcdpara_1})
of spectra in various two, three, and four-nucleon channels which employ controlled approximations, only.
The two extractions of Refs.~\cite{Beane:2012vq,Beane:2013br}~use the same mass for all three light quarks,
set such that \mpill.
The exploratory calculation three-baryon systems in Ref.~\cite{Beane:2009gs}~,
Ref.~\cite{Yamazaki:2012hi}~(510~MeV), Ref.~\cite{Yamazaki:2015asa}~(300~MeV),
Ref.~\cite{Beane:2006mx}~($350\to590~$MeV), and
Ref.~\cite{Orginos:2015aya}~(450~MeV) use the physical value for
the strange-quark mass but degenerate up and down-quark masses corresponding
to the respective $m_\pi$.
At the SU(3) symmetric point, data over a wide range of states, \eg, the H-dibaryon $(\Lambda\Lambda)$, the hyper triton ($^3_\Lambda$H), or hyper Helium-4 ($^4_\Lambda$He), was extracted. No comprehensive theory has been
devised for few-baryon systems comprised of all elements of the $A=1$ octet based on this data and we review the strangeness $s=0$ sector, only.
We begin with the two-body sector where we compare features of the nuclear
bound and scattering systems at physical \mpi~with the LQCD data.
The three and four-nucleon sectors are discussed in parallel, given their correlation.
\paragraph*{Two nucleons  ---}
Eventually, LQCD must produce the two small (relative to $\Lambda_\text{QCD}\sim 4\pi f_\pi\sim\mn\sim 1~$GeV set by $g_s(\Lambda_\text{QCD})\sim 1$) scales characteristic for nuclear physics
in their observed \textit{unnatural} ratio, namely, a momentum scale associated with the poles of the low-energy two-nucleon scattering amplitude, \mbox{$(\anpt)^{-1}\sim\sqrt{\mn B_2(^3S_1)}\equiv\gamma_3\sim 45~$MeV} in the
S-wave spin-triplet and \mbox{$(\anps)^{-1}\sim\gamma_1\sim -8~$MeV} in the S-wave spin-singlet channel, and another related to the typical range of the nuclear interaction $\sim m_\pi^{-1}$.
In addition to this unnaturally large value, $m_\pi/\gamma_3\sim 3$, only the triton binding momentum has to be known for a predictive theory whose breakdown with the nucleon number $A$ is
still unknown.
\begin{table}
\setlength{\tabcolsep}{5pt}
\renewcommand{\arraystretch}{1.4}
\caption{\label{tab.latdata_1}{Few-nucleon data from physical and computational experiments. LQCD uncertainties are upper bounds of the quoted values. Binding energies, $B_A(^{2S+1}L_J)$, are given in MeV, $\gamma_3$ is
the binding momentum of the two-nucleon triplet state, and $B_2^{\pm}$
are the binding energies corresponding to the poles of the effective-range amplitude (see Eq.~(\ref{eq.pole})) parameterized by the central values of
the scattering length $\anpt$ and the effective range $r_3$. Data at \mpill~has been updated\scite{mjs-priv-comm}.}}
\small\centering
\begin{tabular}{l|crcrr}
\hline\hline\rowcolor{white}
 \diagbox[width=10em]{}{\mpi} & $140$ & $300$ & $450$ & $510$ & $806$ \\
 \hline
    $B_{2}(^1S_0)$   & $-$    & $8.5\pm 2.3$  & $12.5\pm 5.0$ & $7.4\pm 1.4$  & $16\pm 5.4$  \\
    $B_2(^3S_1)$     & $2.22$ & $14.5\pm 3.1$ & $14.4\pm 3.2$ &$11.5\pm 1.3$  &$22.6\pm 5.2$ \\
    $B_3(^2S_{1/2})$ & $8.48$ & $21.7\pm 14$  & $-$           &$20.3\pm 4.5$  &$55.8\pm 6.2$ \\
    $B_4(^1S_0)$     & $28.3$ & $47\pm 27$    & $-$           &$43\pm 14$     &$100.6\pm 17$ \\
\hline
    $\frac{\anpt}{r_3}$    & $3.11$  & $-$    & $-3.27\pm14.7$ & $-$    & $2.02\pm 0.64$ \\
    $\frac{m_\pi}{\gamma_3}$ & $3.0$   & $2.4$  & $3.4\pm0.4$    & $4.1$  & $4.2\pm 0.5$   \\
    $B_2^{-}(^1S_0)$       & $-$     & $-$    & $<1$           & $-$    & $12.8$         \\
    $B_2^{-}(^3S_1)$       & $2.21$  &  $-$   & $<1$           & $-$    & $24.0$         \\
    $B_2^{+}(^1S_0)$       & $-$     & $-$    & $12.3$         & $-$    & $25.7$         \\
    $B_2^{+}(^3S_1)$       & $34.8$  & $-$    & $14.2$         & $-$    & $34.5$         \\
\hline\hline
\end{tabular}
\end{table}

Associating scales analogously at heavier \mpi,
this parameter, \ie, in general the scale of the system compared to the range of the interaction,
and here in particular $m_\pi/\gamma$, is still large (Table~\ref{tab.latdata_1},
$6^\text{\tiny th}$~row) and thus suggests an unnatural EFT.
At physical \mpi, the scattering length and effective range provide another pair of scales
that approximate the low-energy two-nucleon spectrum equally well as a consequence of the pion dominating
(ex- or implicitly) the nuclear interaction up to $\sim 100~$MeV, although it cannot account for the emergence
of the unnatural size of the system. Both measures, $m_\pi/\gamma_3$ and $a/r$, are therefore of similar size.
This is also found at \mpif, suggesting a similar analytic structure of the amplitude.
The first fully-dynamical calculation of NN scattering parameters in Ref.~\cite{Beane:2006mx} extracts scattering lengths, only.
These scattering lengths are $\sim m_\pi^{-1}$ (see Table~\ref{tab.latdata_0})
and thus indicate natural NN systems at 350, 490, and 590~MeV pion masses in both spin channels.
Here it is assumed that the interaction range is about the inverse pion mass.
This assumption fails as shown in Ref.~\cite{Beane:2013br}~, where \mpill~does not approximate
the effective interaction range well.
By taking the lattice data for $a$ and $r$ as scales, a more natural theory in which these parameters are of
the same size is implied. Comparing the respective ratio with scales given by the system's
binding momentum and \mpi,~yields a larger ratio.
\begin{table}
\setlength{\tabcolsep}{5pt}
\renewcommand{\arraystretch}{1.4}
\caption{\label{tab.latdata_0}{Two-nucleon scattering lengths, ${}^{2S+1}a_{np}$ (in fm),
and effective ranges, $r_{2S+1}$ (in fm), above the inverse physical and unphysical pion masses (in fm).
The data sources are listed in Table~\ref{tab.lqcdpara_1}.}}
\small\centering
\begin{tabular}{c|rccccr}
\hline\hline\rowcolor{white}
 \diagbox[width=5em]{}{\vspace{5pt}\mpi} & $140$     & $353$         & $450$          & $493$         & $593$        & $806$ \\
 \hline
$\anps$                                  & $-23.75$ & $0.63\pm0.50$ & $20\pm60$   & $0.65\pm0.18$ & $0.0\pm0.5$  & $2.3\pm0.46$ \\
$r_1$                                    & $2.75$   & $-$           & $3.0\pm1.0$ & $-$           & $-$          & $1.1\pm0.14$   \\
$\anpt$                                  & $5.42$   & $0.63\pm0.74$ & $-11\pm55$  & $0.41\pm0.28$ & $-0.2\pm1.3$ & $1.8\pm0.31$ \\
$r_3$                                    & $1.74$   & $-$           & $3.4\pm1.8$ & $-$           & $-$          & $0.91\pm0.14$ \\
\hline
$m_\pi^{-1}$                  & $1.4$    & $0.56$        & $0.44$      & $0.40$        & $0.33$       & $0.25$          \\
\hline\hline
\end{tabular}
\end{table}

\begin{figure}[tb]
\centerline{\includegraphics[width=1. \textwidth]{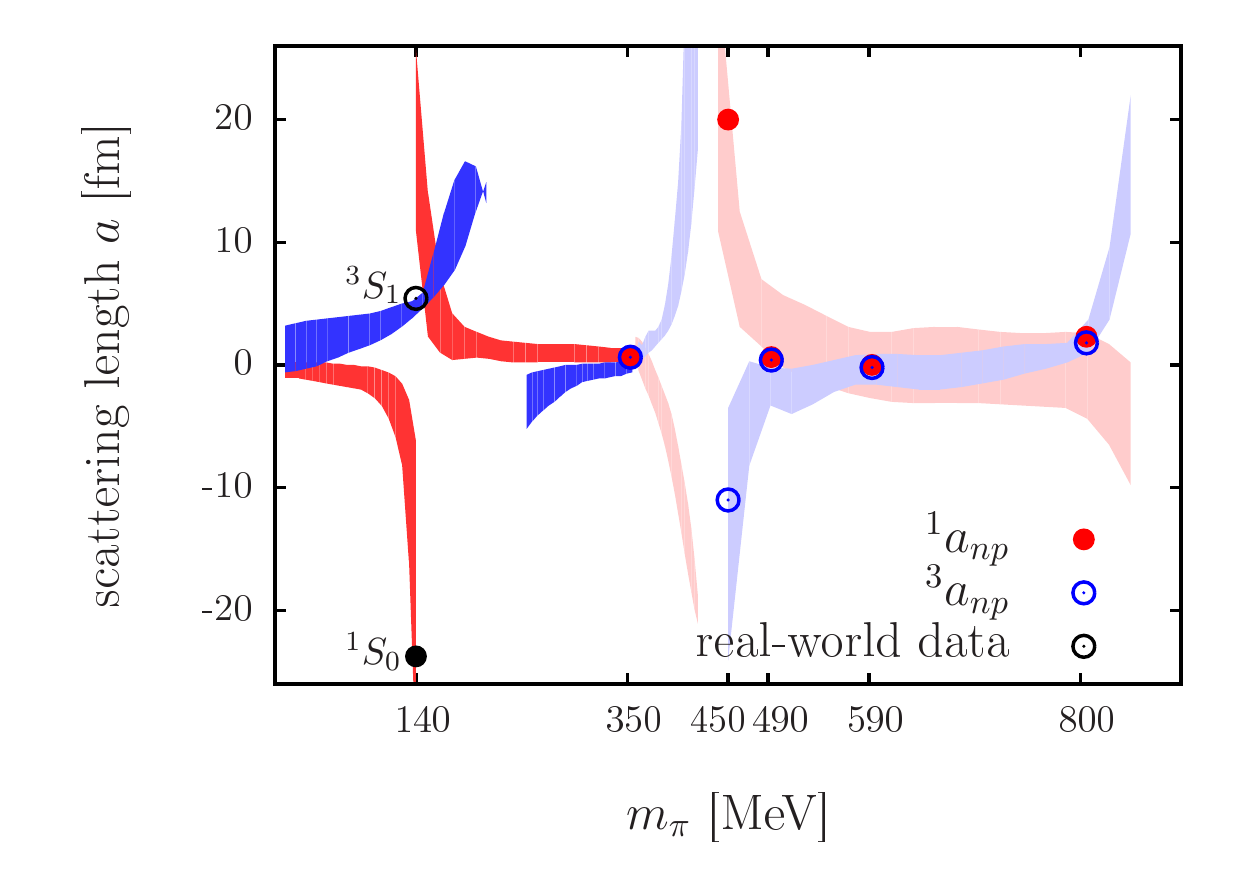}}
\caption{\label{fig.aland} (Color online) Pion-mass dependence of
 the singlet (filled circles) and triplet (empty circle)
 neutron-proton scattering lengths $a$. Allowed regions for $a$ for $m_\pi\lesssim350~$MeV (opaque)
 were derived\scite{Beane:2006mx}with KSW power counting\scite{Beane:2001bc}. 
 The transparent regions are speculations inspired by the limit cycle as observed in the
 running of the three-body contact interaction.}
\end{figure}

A graphical summary of available scattering-length measurements as compiled in Table~\ref{tab.latdata_0}~is
shown in Fig.~\ref{fig.aland}. There, we display the experimental $\anps$ (filled black circle) and
$\anpt$ (empty black circle) together with the lattice data at
larger pion mass ($\anps$: filled red, $\anpt$: empty blue).
The continuation (transparent areas for $m_\pi\gtrsim350~$MeV) of the allowed region of scattering-lengths
(opaque blue for triplet and opaque red for singlet) as extrapolated\scite{Beane:2006mx}using the BBSvK 
power counting of Ref.~\cite{Beane:2001bc}~, is a speculation which
is motivated by the log-periodic running of the three-nucleon momentum-independent
interaction with a regulator parameter (see next paragraph and Ref.~\cite{Bedaque:1999ve})~.
As shown in Fig.~\ref{fig.aland}, the NPLQCD data
indicates a critical pion mass of about $440~$MeV in addition to the one found\scite{Epelbaum:2006jc}at
$198~$MeV at which NN scattering lengths diverge. The limit-cycle assumption of a periodic 
behavior of two-nucleon scattering lengths, with the pion mass instead of a cutoff parameter,
leads to a nuclear interaction at a larger pion mass identical to the one found in nature.
Real nuclei\footnote{Of course, the role of the nucleon mass, and the effective range has to
be taken into account.} could thus be calculated from QCD on the lattice at large pion masses, avoiding
the uncertainty associated with light quark masses. From Fig.~\ref{fig.aland}, one would na\"ively expect
this copy of the real nuclear two-nucleon world at about $850~$MeV.

We turn the discussion to data sets which include effective ranges.
Scales are related to the analytic structure of an amplitude, \ie,
the position of \textit{a} pole and the radius of convergence of some expansion around that pole. In case of the effective range expansion at physical \mpi, $\gamma\sim a^{-1}$ and
$r\sim m_\pi^{-1}$, respectively. In that sense, $\gamma\sim\sqrt{B_2\mn}\sim k_\pm(a,r)$ (see poles of Eq.~(\ref{eq.pole}) below) defines a scale even at \mpill. But $\gamma\sim\sqrt{B_2\mn}\gg k_\pm(a,r=0)$
and $m_\pi/\gamma\gg a/r$ (Table~\ref{tab.latdata_1}) imply additional non-analyticities within
an \mpi~radius around $\gamma$ beside the existing bound-state pole.

\begin{figure}[tb]
\centerline{\includegraphics[width=1. \textwidth]{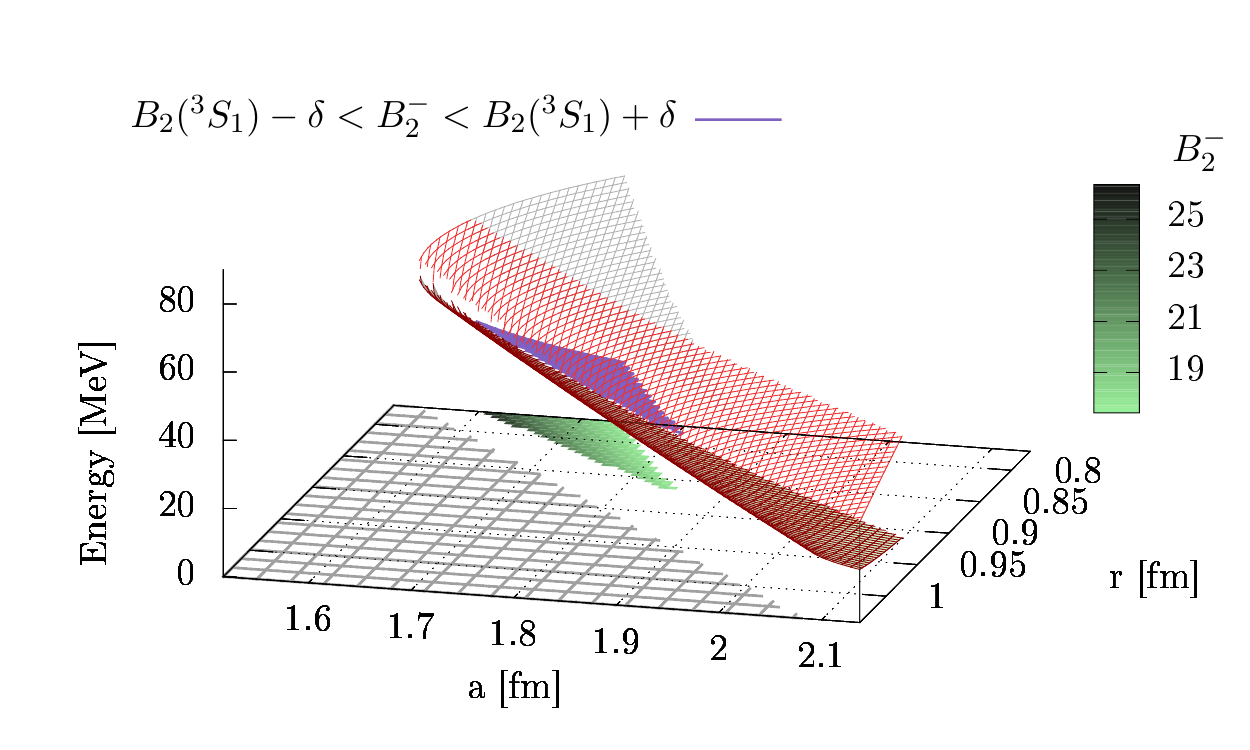}}
\caption{\label{fig.2bdyland} (Color online) Dependence of two-nucleon binding energies as poles of the ERT scattering amplitude parameterized\scite{Beane:2013br}at \mpill~
via \mbox{$\anps=2.33\pm 0.46~$fm}, \mbox{$r({}^1S_0)=1.13\pm 0.14~$fm}, and \mbox{$\anpt=1.83\pm 0.31~$fm}, \mbox{$r(^3S_1)=0.91\pm 0.14~$fm}. $r-a$ pairs from the gray lattice area in the $xy$ plane allow virtual or resonant states. $r-a$ pairs defining an energy $B_2^-$~(purple patch of the surface) consistent with the value extracted with a ``bound-state''
source (see $B_2(^3S_1)$ in Table~\ref{tab.latdata_1}) \textit{and} a gap to $B_2^+\gtrsim 30~$MeV (light gray patch)
are marked in green.
Binding energies with $B_2^+-B_2^-\lesssim 30~$MeV are shown as red surfaces. The two surfaces touch at a line which
marks the double pole of the amplitude.}
\end{figure}
To parameterize the scattering amplitude using effective-range theory (ERT) as
\begin{equation}\label{eq.pole}
T(k)=\frac{4\pi}{\mn}\frac{1}{k\cot\delta-ik}=\frac{4\pi}{\mn}\frac{1}{-\frac{1}{a}+\frac{r}{2}k^2+\mathcal{O}(k^4)-ik}
\end{equation}
is justified through a pole consistent with the non-perturbative binding
energy (compare the $1^\text{\tiny st}$($2^\text{\tiny nd}$) and $7^\text{\tiny th}$($8^\text{\tiny th}$)
row in Table.~\ref{tab.latdata_1}).
The amplitude Eq.~(\ref{eq.pole}) has two poles at momenta $k_\pm=\frac{i}{r}(1\pm\sqrt{1-2r/a})$, \ie, possibly two bound states.
At physical \mpi, $k_-^2/\mn\sim2.2~$MeV, the binding energy of the deuteron, while $k_+$ is beyond the range of validity of the ERT and therefore not in disagreement with the missing experimental evidence
for such a deep state. At \mpill, $2r/a\sim 1$ in both S-wave channels
(see also discussion in Sec.~\ref{sec.left}) which implies relatively closely spaced bound states.
Considering the uncertainty in the lattice extraction of the scattering length and
effective range yields via Eq.~(\ref{eq.pole}) the following scenarios of the two-nucleon spectrum:
\textit{(i)} two shallow states for $r/a=(2+\epsilon)^{-1}$ (dark and light red areas on the energy surface in
Fig.~\ref{fig.2bdyland}), \textit{(ii)} one shallow state
and a spurious deep state if $r/a\to 0.3$, \ie, the analog to physical \mpi~(gray energy surface and its green projection onto the $a-r$ plane),
and \textit{(iii)} \textit{no} bound states but resonances for $r/a\to 0.6$ (hatched gray area in $a-r$ plane). The limits correspond to the smallest and largest ratios consistent with the uncertainties.

Scenarios \textit{(i)} and \textit{(ii)} are also consistent with the spectral measurement of $\eodim=22.3\pm 5~$MeV albeit only one bound state has been isolated. To discriminate a second, almost degenerate state is at present impossible
as the individual extraction of energy levels from the effective-mass plots suffers from significant statistical noise. A ratio measurement between a singlet and a triplet two-nucleon bound state correlator will be less affected
by that noise and could hint towards a second bound state.
In Sec.~\ref{sec.left}, we comment on the implication of this peculiar analytic structure for the construction of an EFT.
\paragraph*{Three and four nucleons  ---}
For heavier systems, LQCD provides, at present, only data on
bound states, until numerical precision will suffice to identify
signals corresponding to two-fragment continuum states thus enabling the calculation of, \eg,
nucleon-deuteron phase shifts via a generalization\scite{Bour:2011ef}of Eq.~(\ref{eq.luescher}).

At physical \mpi, the existence of bound states in the three-nucleon system is a universal consequence\scite{Bedaque:1999ve}of the proximity of the two-nucleon system to the unitary limit $|\anpst|\to\infty$.
The fact that there is only one such state in the $J^\pi=\frac{1}{2}^+$ channel and its exact binding energy are then peculiar to the three-nucleon system and not fixed by the two-body scattering length and
effective range. Furthermore, there is no physical bound state in the $\frac{3}{2}^+$ channel.
In contrast to this independence of $\eotri$ of the two-body $a$'s and $r$'s,
the three and four-nucleon spectra are intertwined, and the emergence of two four-body states
with every three-body state, as known from
unitary bosonic systems\scite{Hammer:2006ct}, seems to generalize to the nuclear
problem\scite{Platter:2004zs,Kirscher:2009aj}.
Whether or not the Coulomb repulsion transforms one of the two bound $J^\pi=0^+$ states to a shallow resonance and the other to the $\alpha$-particle will be a test for \eftnopi~and its treatment of
electromagnetism in few-body systems.

The LQCD calculations at higher \mpi~identify single bound states in the triton and $\alpha$ channels, too. There is no data on other three or four-nucleon channels.
In the strangeness $s=-1$ sector, the NPLQCD
collaboration\scite{Beane:2012vq}finds a bound $\frac{3}{2}^+$ hyper triton.
At the considered SU(3) symmetric point, this state hints
at a corresponding bound $s=0$ quartet state. For $s=-2$, a second
four-body $0^+$ bound state is found, reminiscent of the aforementioned
universal tetramer pair associated with each three-body state in the
two-body unitary limit.
Given the degenerate ${}^3$H and $^3_\Lambda$H $\frac{1}{2}^+$ three-body, and $\alpha$ and $^{\;\;\;4}_{\Lambda\Lambda}$He four-body states, and the above conjectured two-level structure of the two-nucleon system, LQCD investigations
of possible $\frac{3}{2}^+$ three and additional $0^+$ four-nucleon bound states could reveal striking differences of few-nucleon systems at larger pion masses.
\begin{table}
\setlength{\tabcolsep}{5pt}
\renewcommand{\arraystretch}{1.4}
\caption{\label{tab.latdata_3}{Binding energies (in MeV) of the shallowest two, three, and four-nucleon state relative to its lowest break-up threshold, \ie,
$\Delta_{21}\equiv B_2(^1S_0)$ ($B_2(^3S_1)$ at $m_\pi\sim140$~MeV), $\Delta_{32}\equiv B_3(^2S_{1/2})-B_2(^3S_1)$, and $\Delta_{43}\equiv B_4(^1S_0)-B_3(^2S_{1/2})$.
Uncertainties (insignificant at $m_\pi\sim140$~MeV) were combined in quadrature. References to LQCD data at unphysical \mpi~are the same as in Table~\ref{tab.lqcdpara_1}.}}
\small\centering
\begin{tabular}{c|cccc}
\hline\hline\rowcolor{white}
 \mpi  & $\Delta_{21}$ & $\Delta_{32}$ & $\Delta_{43}$ & $\Delta_{43}/\Delta_{32}$ \\
 \hline
$140$ & $2.2$         & $6.3$ & $20$ & $3.2$ \\
$300$ & $8.5\pm 2.3$ & $7.2\pm 17$ & $25\pm 41$ & $3.5\pm 14$ \\
$510$ & $7.4\pm 1.4$ & $8.8\pm 5.8$ & $23\pm 19$ & $2.6\pm 3.8$ \\
$806$ & $16\pm 5.4$  & $33\pm 11$ & $44\pm 24$ & $1.4\pm 1.2$ \\
\hline\hline
\end{tabular}
\end{table}

Beside the sheer existence, the relation of binding energies,
\ie, thresholds, and binding energies per nucleon in nuclei of
different $A$ are relevant for an effective description of heavier systems.
At physical \mpi, the triton is bound by $\sim 6.5~$MeV relative
to the deuteron-neutron threshold and interpreted as shallow, corresponding to an Efimov state.
The $\alpha-$particle is bound by $\sim 20~$MeV relative to triton-proton and considered a universal feature.
Some hints to whether or not this interplay between two, three, and four-nucleon states persists at higher \mpi~are found in the relative threshold positions, compiled in Table~\ref{tab.latdata_3}.
For all $m_\pi>140~$MeV, the triton is closer to the deuteron-nucleon threshold than the $\alpha-$particle is to the noninteracting triton-proton system. The three-nucleon ($\Delta_{32}$) and two-nucleon break-up energies are of the
same order and decrease from 806 to 510~MeV. At $m_\pi\sim300$~MeV, a bound or an unbound three-nucleon system are within uncertainty limits.

The gap between the measured bound state energy of $\alpha$ and its lowest break-up threshold $\Delta_{43}$ is larger than $\Delta_{32}$ but of the same size or smaller than the scale set by $\eodim$.
The ratio $\Delta_{43}/\Delta_{32}$ decreasing with \mpi~and $\Delta_{43}$ increasing simultaneously, is reminiscent of the scattering-length dependence of the two and three-body systems for $a\to\infty$.
With decreasing \mpi, the triton approaches threshold.
The discussion below on the neutron-deuteron scattering length suggests a diverging three-nucleon amplitude at zero energy and thereby the analog of a limit cycle in the four-nucleon, three-body deuteron-nucleon-nucleon system.

We summarize this section at the beginning of Sec.~\ref{sec.left} which reviews a theory trying to describe the data consistently.
\section{Nuclear theories}
The interpretation of the above results as a data base for a theoretical analysis
is analogous to the way early experiments on nuclei initiated theoretical nuclear physics.
A theoretical analysis of lattice measurements is justified if computational resources are not
expected to be available in the near future for the nuclear properties of interest.
Two approaches to a systematic understanding of nuclear lattice data are available.
In effect, they generalize
the concept presented in Sec.~\ref{sec.chi2nopi} by matching a nuclear contact theory to
LQCD amplitudes and the application of the ensuing theory to few-nucleon systems.
The two methods differ in the matching condition \textit{and} the contact theory. How exactly those
differences lead to inconsistent postdictions is not known\footnote{For recent work on the assessment of the sensitivity of few-hadron LQCD results on the source structure see Ref.~\cite{aok-priv-comm}~.}.
Thus, we deem a brief summary of their respective technique and basic assumptions as useful. 
\subsection{Matching wave functions}\label{sec.hal}
Analogous to an approach taken for the description of Kaon decay\scite{Lin:2001ek}, a connection
between QCD four-point correlation functions and non-relativistic nuclear potentials was made
(the original work is Ref.~\cite{Ishii:2006ec}~, for a review see Ref.~\cite{Aoki:2013tba}~).
While matching as reviewed between \ce~and experiment or \eftnopi~identified amplitudes, here,
two theories are matched through a set of wave functions.
The underlying theory QCD defines a wave function which can be extracted with the lattice technology.
This function satisfies a non-relativistic Schr\"odinger equation if the
potential is chosen appropriately. The unknown is thus the potential while the wave function is input.

More specifically, a relativistic Nambu-Bethe-Salpeter wave function
(introduced in Ref.~\cite{Chu199131}~for single hadrons) is extracted
with a correlation function as in Eq.~(\ref{eq.efmf}).
Ref.~\cite{Ishii:2006ec}~, in particular, uses a so-called wall source for
\mbox{$\overline{N}_\text{source}(\lbrace\bs{\rho}_i^\text{  source}\rbrace_{i=1,\ldots,A-1},t_0=0)$}, \ie,
the local quark fields in the interpolating operators are projected onto zero momentum.
The source specifically creates a two-nucleon state with defined parity,
isospin $I$ and total angular momentum $J$. At the sink, the two-nucleons are annihilated with the same quantum numbers but at a selected distance $\bs{x}$ apart:
\begin{equation}\label{eq.NBS}
C(\bs{x},t-t_0)=\left\langle N_{1,I,J}(\bs{r},t)N_{2,I,J}(\bs{r}+\bs{x},t)\overline{N}_{1,I,J}(t_0)\overline{N}_{2,I,J}(t_0)\right\rangle\;\;.
\end{equation}
We iterate the aforementioned ambiguities in the calculation of this correlation function: the structure of the interpolating fields at the quark \textit{and} hadron level.

Two methods have been used to extract a two-nucleon potential from the QCD Greens function $C(\bs{x},t-t_0)$ in Eq.~(\ref{eq.NBS}).
The standard used to predict masses and ground-state energies of composite objects in LQCD uses the spectral decomposition
of the correlation function as given above in Eq.~(\ref{eq.efmf}) which reads in the case of interest here,
\begin{eqnarray}\label{eq.sat}
C(\bs{x},t-t_0)&=&\sum_{n}Z_n\cdot\Psi(\bs{x};\bs{k}_n,I,J)\cdot e^{-E_n(t-t_0)}\nonumber\\
&\to&Z(\bs{k}_0,I,J)\cdot\Psi(\bs{x};\bs{k}_0,I,J)\cdot e^{-2\sqrt{\bs{k}^2+\mn^2}(t-t_0)}\;\;.
\end{eqnarray}
For large enough $t-t_0$ the ground state can be isolated if the statistical noise is sufficiently reduced.
To reduce this noise is a challenge for for the identification of energies from effective-mass plots, in general,
and can be achieved through an increased \mpi.
It is problematic for the extraction of wave functions, in particular,
as shown with a toy model in Ref.~\cite{Birse:2012ph}~, which exemplifies how admixtures to the wave function
$\Psi(\bs{x};\bs{k}_0,I,J)$ stemming presumably from other states in the spectrum
modify a potential derived from it.
That model, specifically, found the hard core, relative to the outer tail,
of the potential to be highly sensitive to the coupling to other channels.
Thus the amount of ground state saturation in
Eq.~(\ref{eq.sat}) controls the range of applicability of the potential~---~
the stronger the admixture, the smaller the energy at which the potential becomes useless.

An alternative extraction method which overcomes the saturation issue was devised in Ref.~\cite{HALQCD:2012aa}~
(see Ref.~\cite{Charron:2013paa}~for a comparison).
It cleverly exploits the fact that each wave function in the spectral sum Eq.~(\ref{eq.sat}) satisfies a
Schr\"odinger equation. Therefore, one can use a rescaled correlation function
$R(\bs{x},t)\equiv C(\bs{x},t)/(e^{-\mn t})^2$
to define a potential $U$ which, in general, acts between two-nucleon channels and thus shall be, like $C$, understood as a matrix, via
\begin{equation}\label{eq.imex}
\left\lbrace -H_0-\frac{\partial}{\partial t}+\frac{1}{4\mn}\frac{\partial^2}{\partial t^2}\right\rbrace R_{\alpha}(\bs{x},t)=
\sum_{\alpha'}\int d^3yU_{\alpha\alpha'}(\bs{x},\bs{y})R_{\alpha'}(\bs{y},t)\;\;.
\end{equation}
The same potential can be constructed from the ground-state wave function obtained via Eq.~(\ref{eq.sat}) from the stationary
Schr\"odinger equation
\begin{equation}\label{eq.stex}
\left\lbrace -H_0+\frac{\bs{p}^2}{\mn}\right\rbrace \Psi_{\alpha}(\bs{x})=
\sum_{\alpha'}\int d^3yU_{\alpha\alpha'}(\bs{x},\bs{y})\Psi_{\alpha'}(\bs{y})\;\;.
\end{equation}
A non-local, velocity-dependent ansatz
\begin{equation}\label{eq.nonlocp}
U(\bs{x},\bs{y})=V(\bs{x},\bs{\nabla})\delta^{(3)}(\bs{x}-\bs{y})=V_0+V_\sigma\bs{\sigma_1}\cdot\bs{\sigma_2}+V_TS_{12}+V_\text{LS}\bs{L}\cdot\bs{S}+\mathcal{O}(\bs{\nabla}^2)
\end{equation}
was chosen\scite{Ishii:2006ec}for the nuclear potential in a given channel. The expansion shown on the right-hand side
defines \mbox{$\vert\bs{x}\vert$-dependent} coefficient functions which are constructed order by order inverting either
Eq.~(\ref{eq.imex}) or Eq.~(\ref{eq.stex}).

Nuclear observables in the strangeness $s=0$ sector predicted with potentials as defined above relying on unquenched lattice QCD correlation functions (see Table~\ref{tab.lqcdpara_1})
include the scattering length in the two-nucleon $^1S_0$ channel with c.m. momentum $k$ derived in the limit
\begin{equation}\label{eq.halsc}
\anps(m_\pi=701~\text{MeV})=\lim\limits_{k\to 0}k^{-1}\tan\delta(k)=1.6\pm1.1~\text{fm}\;\;,
\end{equation}
which was obtained from the central part in Eq.~(\ref{eq.nonlocp}). Tensor and spin-orbit potentials have also been derived\scite{Murano:2013xxa}at \mpi~up to 1.1~GeV
and applied in variational calculations\scite{Inoue:2011ai}. Neither two nor three-nucleon bound states but indications
for a shallow four-nucleon state, whose binding energy increases from $\sim 0.8~$MeV to $\sim 5.1~$MeV with \mpi~decreasing over the considered range, have been found.
Employing the same potentials via the Brueckner-Hartree-Fock method to $^{16}$O and $^{40}$Ca nuclei yields those nuclei bound consistent with the $5.1~$MeV for the $\alpha$-particle.

Future work deriving a three-nucleon potential, as pioneered at $m_\pi\sim1.1$~GeV in Ref.~\cite{Doi:2011gq}~,
will tell whether or not the above expansion of the nuclear interaction
receives a significant contribution from such a force relative to higher-orders of the velocity expansion of the two-nucleon potential. In the following section~\ref{sec.left}, we will present an analysis of
the role of three-nucleon force and its \mpi~dependence in an EFT framework.

\subsection{Matching correlation functions}\label{sec.left}
A theory for few-nucleon systems which explains the LQCD data at $m_\pi\sim510$~MeV and
\mpill~(for lattice data see Refs.~\cite{Yamazaki:2012hi,Beane:2012vq}~,
and for the EFT Refs.~\cite{Barnea:2013uqa,Kirscher:2015yda}~)
resembles the effective-field-theory approach to low-energy
few-nucleon systems in the physical world\scite{vanKolck:1998bw,Chen:1999tn,Kaplan:1998we,Kaplan:1998tg}.
Compared to the method introduced in the previous section,
it matches to QCD via observable binding energies instead
of wave functions. The matching conditions, as part of the
definition of an EFT, replace real-world data with LQCD
predictions for the input~---~we are exploring the domain
of the transparent sheets in Fig.~\ref{fig.theoland}.
\begin{figure}[tb]
\centerline{\includegraphics[width=1. \textwidth]{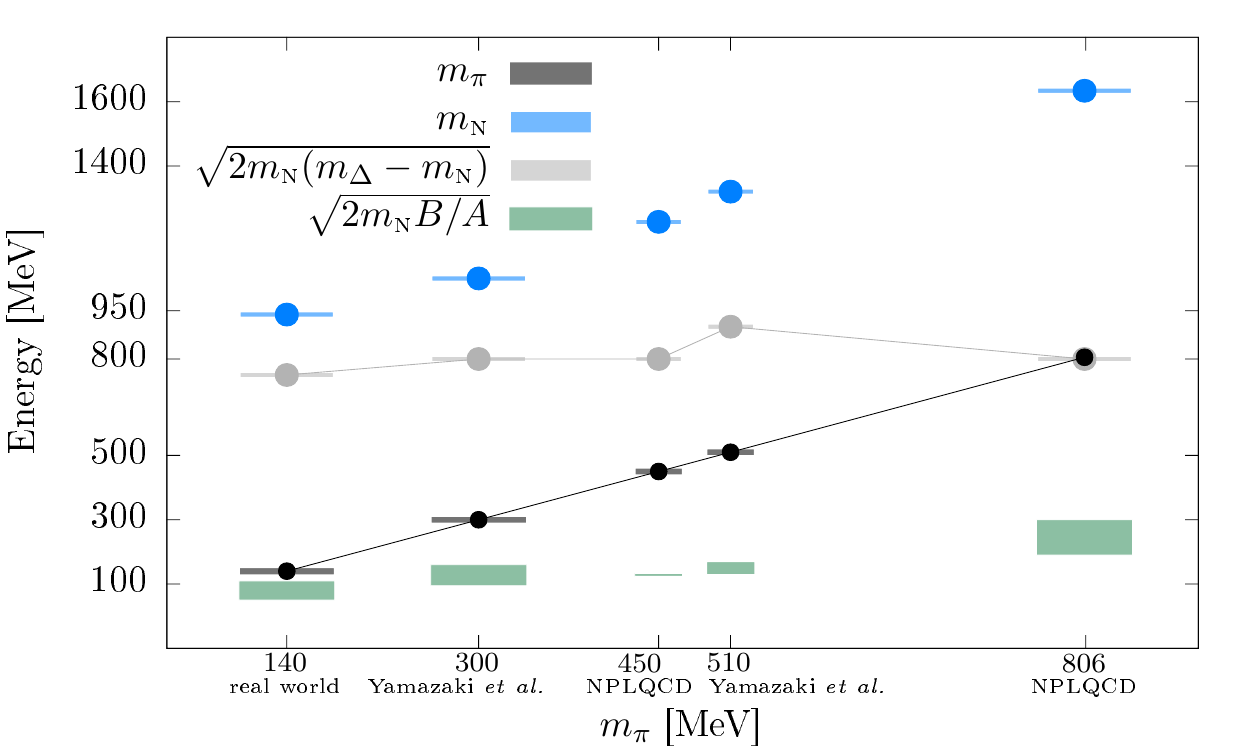}}
\caption{\label{fig.scales} (Color online) Pion-mass dependence of momentum scales relevant for nuclear low-energy physics. The green-shaded boxes represent the range of binding momenta $\sqrt{2\mn B/A}$ for
nuclei with $A=2,3,4$ and binding energy $B$ as given in Table~\ref{tab.latdata_1}. The pion-mass (black) and $\Delta$ (gray, see footnote~\ref{ftn.delta}) scales become equal for \mpill.}
\end{figure}

Nucleons as eigenstates of QCD are canonically defined as an isospin doublet belonging to the lowest-mass baryon SU(3) octet. The first amendment to the most general SU(3) invariant Lagrangian, constructed solely from this octet,
couples it, first, to the lowest-mass meson octet and second, to the lowest-mass baryon decuplet.
A comparison between \textit{(i)} the nucleon mass $\mn$, \textit{(ii)} a scale associated with the excitation of a nucleon to a $\Delta$\scite{Savage:1996tb}, $\sqrt{2\mn(m_\Delta-\mn)}$,
\textit{(iii)} the mass of the pion \mpi, and \textit{(iv)} the binding momenta of an $A$-body nucleus, $\sqrt{2\mn B/A}$ gives an indication whether an approximation of those couplings,
specifically, pion-nucleon, $\Delta$-nucleon, and $\Delta$-pion-nucleon, with contact interactions amongst nucleons might be useful over some energy range for nuclear amplitudes.

These scales are compiled in Fig.~\ref{fig.scales} for the different pion masses and led to
the ansatz\scite{Barnea:2013uqa}of a nuclear contact theory analog to the established
\eftnopi~at physical \mpi.
In essence, this analogy utilizes: \textit{(i)} the much smaller typical binding momenta of nuclei
relative to the nucleon mass to justify a non-relativistic treatment implying small-momentum Lorentz symmetry;
\textit{(ii)} with the typical momenta in bound
systems~---~identified with $Q$~---~up to $A=4$ being smaller
than the lightest meson and lowest baryon excitation\footnote{\label{ftn.delta}The scale in Fig.~\ref{fig.scales}, in contrast,
considers the effect of intermediate states under natural assumptions for the LECs. Under these assumptions, \mpi~remains the lowest threshold setting the convergence rate of the theory.}
$\mn-m_\Delta$~---~collectively denoted $M$~---~
an (iso)spin $1/2$ nucleon field ($N=(n,p)$) suffices as sole degree of freedom;
\textit{(iii)} the renormalized \textit{two-body} LECs are assumed to be of
order $C_{2n}=\frac{4\pi}{\mn Q(QM)^n}$ and hence only iterations of the two $n=0$ zero-derivative
interactions are of the same order. The infinite iteration introduces the two-nucleon bound state
poles; \mbox{\textit{(iv)} the}
three-nucleon momentum-independent interaction in the $^2S_{1/2}$ (triton)
channel is considered at the same order as the momentum-independent two-nucleon
interactions in the $^1S_0$ and $^3S_1$ channels.
This unnatural enhancement \textit{and} the non-perturbative treatment
assumes that the leading three-body interaction follows a limit cycle
analogous to physical \mpi~in Ref.~\cite{Bedaque:1999ve}~.
This approach was demanded by the cutoff dependence of the triton ground
state with only the leading-order two-nucleon interaction.
Two observations support the underlying assumption of a limit cycle:
first, the parabolic increase of the triton ground state energy as a function of the cutoff
in the absence of a three-body force,
and second, the appearance of an additional state at some critical cutoff value.
\begin{figure}[tb]
\centerline{\includegraphics[width=1. \textwidth]{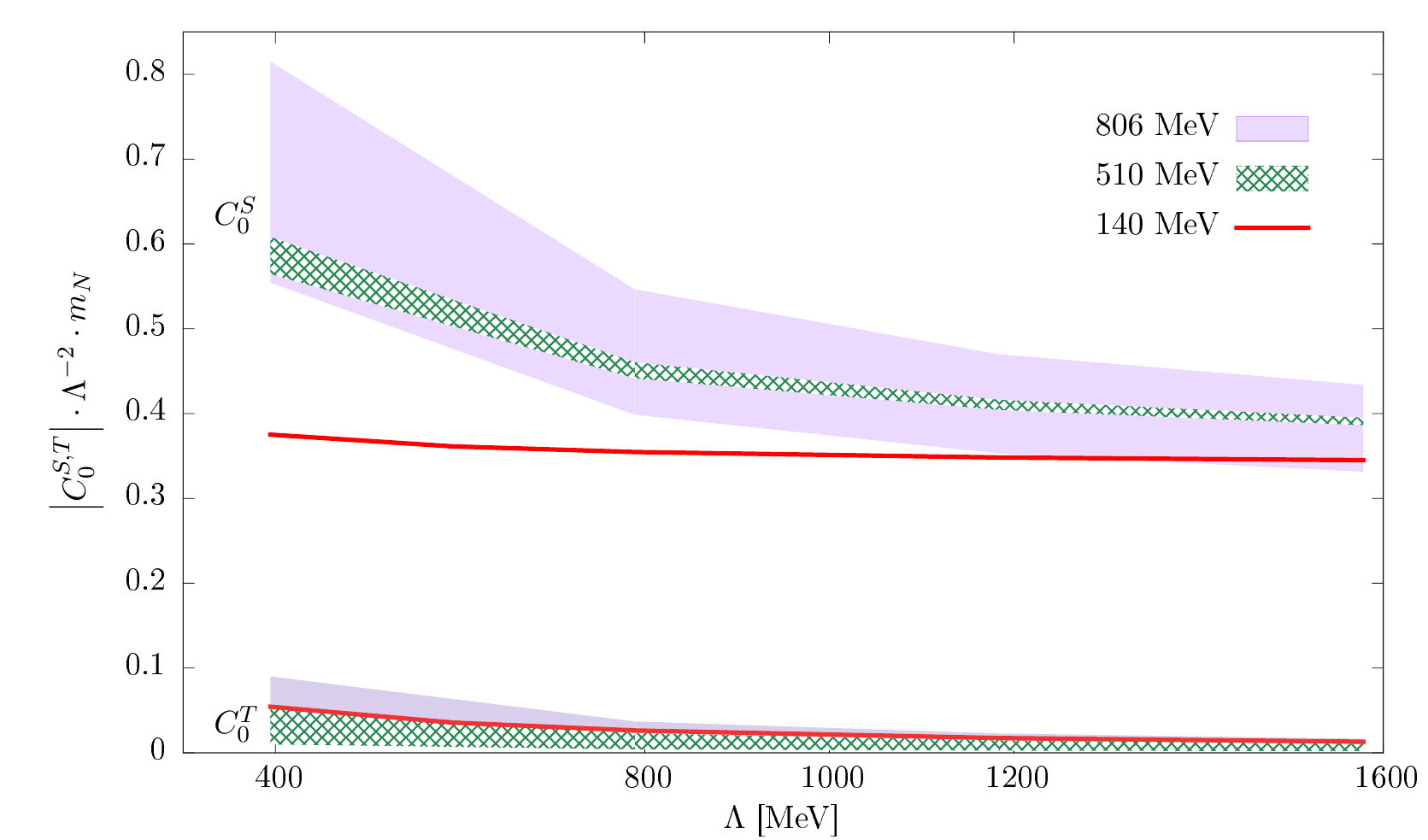}}
\caption{\label{fig.lecs} (Color online) Dimensionless couplings of the lowest-order two-body operators of \eftnopi~at $m_\pi\sim 510~$MeV (hatched), \mpill~(solid), and physical \mpi~(red solid line).
The upper (lower) bands show the regulator dependence of the SU(4) symmetric (asymmetric) LEC $C_0^{S(T)}$ defined in Eq.~(\ref{eq.lagr}). The band widths represent uncertainty in the data used for the calibration.}
\end{figure}

The Lagrange density\footnote{Space-time and field coordinates are rescaled such that $\partial_t$ and $\bs{\nabla}^2/(2\mn)$ are of the same order as in NRQCD (see comment below).}
defining the theory together with the size estimates for the LECs in an operator basis that splits SU(4) symmetric ($C_0^S$) and asymmetric ($C_0^T$) components reads
\begin{equation}\label{eq.lagr}
\mathcal{L}=N^\dagger\left(i\partial_t+\bs{\nabla}^2/(2\mn)\right)N-\frac{1}{2}\left(C_0^S\left(N^T N\right)^2+C_0^T\left(N^T\bs{\sigma}N\right)^2\right)+D_1\left(N^T N\right)^3\;\;.
\end{equation}
The EFT and power-counting scheme implied by \textit{(i-iv)} was
implemented\scite{Barnea:2013uqa,Kirscher:2015yda} as a \textit{cutoff} EFT in coordinate space. The employed renormalization comprised \mbox{\textit{(i)} the} regularization of contact interactions
with Gaussian functions on the relative coordinate $\delta(\bs{r})\to \frac{\Lambda^3}{8\pi^{3/2}}e^{-\frac{\Lambda^2}{4}\bs{r}^2}$, and \textit{(ii)} the calibration of the $\Lambda$-dependent LECs to $\eonn,~\eodim,~$and $\eotri$.
The necessary calculations numerically solved the appropriate two and three-body Schr\"odinger equations.

An approximate Wigner SU(4) symmetry of nuclear interactions at \mpill~has been noted in light of the degenerate (within lattice uncertainty) two-nucleon $^1S_0$ and $^3S_1$ ground-state
energies in Ref.~\cite{Beane:2012vq}~and scattering length to effective range ratios
$a/r\sim 2$ in Ref.~\cite{Beane:2013br}~in the respective channels.
This independence of the nuclear force \wrt~SU(4) spin-isospin transformations\footnote{$\mu,\nu\in\lbrace1,2,3,4\rbrace$, $\alpha$ parameterizes an infinitesimal SU(4) transformation with SU(2) generators $\sigma_\mu(\tau_\nu)=\lbrace 1,\bs{\sigma}(\bs{\tau})\rbrace$
acting on (iso)spin degrees of freedom.}
$\delta N=i\alpha_{\mu\nu}\sigma_\mu\tau_\nu$
translated into the LECs as shown in Fig.~\ref{fig.lecs}.
While the relatively small SU(4) asymmetric LEC $C_0^T$ can be understood from the large scattering lengths in both channels\scite{Mehen:1999qs}at physical \mpi~(lower red solid line in Fig.~\ref{fig.lecs}), the
degeneracy $\anps\sim\anpt\sim 2~$fm is explicit at \mpill. At physical \mpi, SU(4) invariance results from being close to the unitary limit, while at larger \mpi, it seems to emerge as a unique feature of QCD!
\begin{table}
\renewcommand{\arraystretch}{1.1}
\caption{\label{tab.eftres}{Leading-order \eftnopi~post ($m_\pi\sim140$~MeV, $\eotet$) and
predictions\scite{Barnea:2013uqa,Kirscher:2015yda}for the quartet and doublet neutron-deuteron scattering
lengths $\andq$ and $\andd$ and for the $\alpha$-particle binding energy $\eotet$ at three pion masses.}}
\small\centering
\begin{tabular}{l|ccc}
\hline\hline
\mpi~[MeV]         & $140$          & $510$          & $806$        \\ 
\hline
$\andq~$[fm]       &  $5.5\pm 1.3$  & $2.3\pm 1.3$   & $1.6\pm 1.3$ \\
$\andd~$[fm]       &  $0.61\pm 0.50$ & $2.2\pm 2.1$ & $0.62\pm 1.0$ \\
\hline
$\eotet~$[MeV]     & $24.9\pm 4.3$ & $35\pm 22$ & $94\pm 45$\\
\hline\hline
\end{tabular}
\end{table}

\begin{figure}[tb]
\centerline{\includegraphics[width=1. \textwidth]{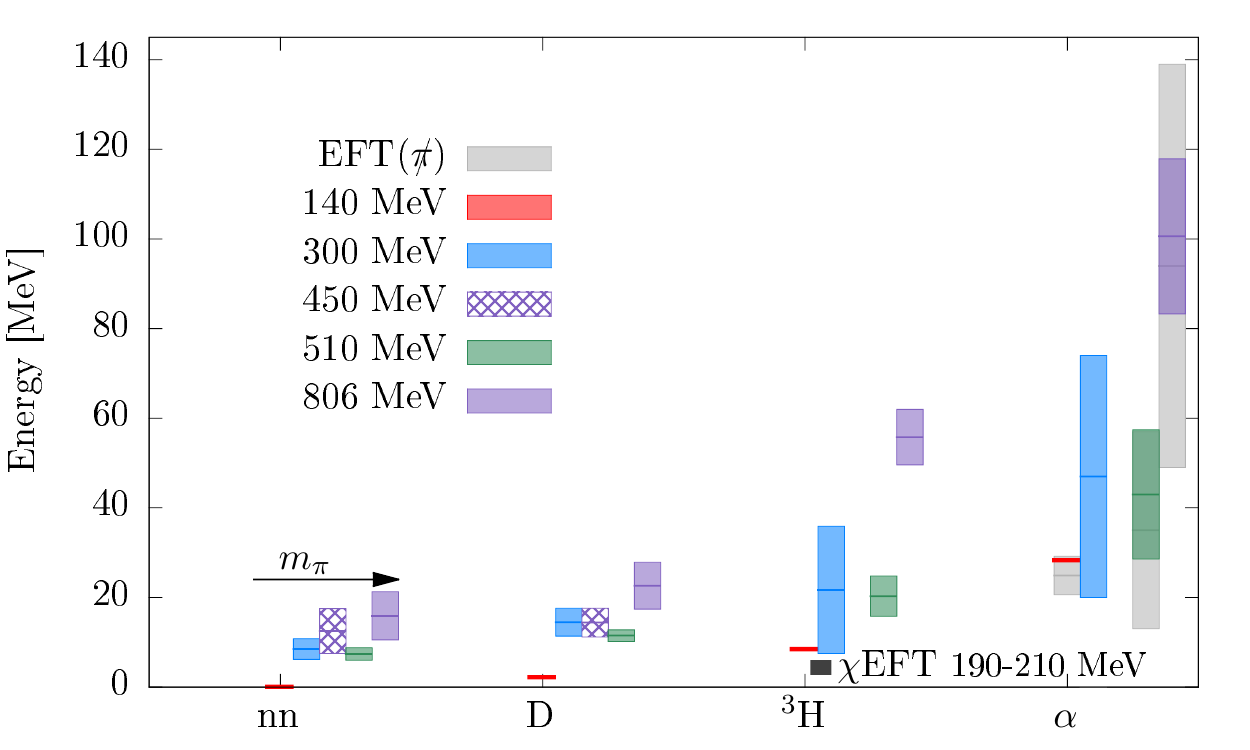}}
\caption{\label{fig.spectrum} (Color online) Lowest energy levels in the $^1S_0$ and $^3S_1$ two-nucleon, the $J^\pi=\frac{1}{2}^+$ three-nucleon triton and $0^+$ $\alpha$ channel.
For each system, the physical (red) value is compared with LQCD measurements at $m_\pi\sim300$ (blue), 450 (hatched purple), 510 (green), and 806~MeV (purple). Postdictions for $\eotet$ of \eftnopi~are in gray.
The black box represents \ce~predictions\scite{Epelbaum:2006jc}and
its height the considered \mpi~range of (190,210)~MeV. The other box heights resemble the lattice uncertainties.}
\end{figure}

The power counting is justified via RG invariance. Namely, for a \textit{cutoff} EFT, predictions
must converge at every order if the RG-flow parameter $\Lambda\to\infty$. Furthermore, cutoff dependence at
a given order can be eliminated at some higher, not necessarily the next, order, where an LEC with the right 
scaling counters the dominating $\Lambda$ dependence which is not eliminated by the lower-order LECs\footnote{We understand $\Lambda$ as a placeholder for any cutoff introduced to regulate the theory.
It is common, \eg, in solving the three-body problem with the dibaryon formalism to use a combination of
cutoff and dimensional regularization. In that case, the limit $\Lambda\to\infty$ must be taken in both
regularization schemes.}.
At present, leading-order predictions with $\Lambda=2-8~\text{fm}^{-1}$ are
available with no sign of inconsistencies in the power counting.
Specifically, the $\alpha$ binding energy $\eotet$, the nucleon-deuteron scattering lengths in doublet $\andd$~and quartet $\andq$ were analyzed in that light.
The former as a bound-state observable that can be benchmarked with the
lattice data, and the latter as scattering properties which constitute predictions,
LQCD will\scite{Beane:2012vq}measure given the resources to increase its numerical accuracy.
The predicted observables are compiled in Table~\ref{tab.eftres} and also included as gray columns in Fig.~\ref{fig.spectrum} for $\eotet$. In contrast to physical \mpi, where the uncertainty of real-world experiments is
insignificant relative to the absolute observed value,
$\eftnopi$~applied to lattice data has to propagate the uncertainty
in the amplitudes used to renormalize the theory, namely $\eonn,~\eodim$, and $\eotri$. To justify the power 
counting, $\Lambda$-variation suffices but was
shown\scite{Barnea:2013uqa,Kirscher:2015yda}to converge slowly.
Calculations within the dibaryon formalism, which is more flexible in its
regulator, could probe the sensitivity of nuclear systems at high quark masses
to short-distance structure more comprehensively.

A different power counting altogether is required for the case of degenerate \mbox{($r/a=1/2$)} or closely-spaced\linebreak \mbox{($r/a=1/2\pm\epsilon$)} two-nucleon bound states in the same spin channel. In the case of \mbox{$r/a=1/2$}, the
effective-range amplitude Eq.~(\ref{eq.pole})~has a double pole which cannot result from an iteration of momentum-independent terms as in Eq.~(\ref{eq.lagr}). The breakdown of a contact as an approximation of a Yukawa
theory\footnote{This is to be understood as any mechanism that produces a Yukawa potential, \eg, the coupling of the nucleon to a scalar field.}
with the latter sustaining a shallow and an excited state was demonstrated in Ref.~\cite{Luke:1996hj}~.
The similarities indicated in Ref.~\cite{Luke:1996hj}~of the presumed two-nucleon LQCD spectrum with more than one shallow state to non-relativistic QCD (NRQCD)
become more detailed in the description of heavy quarkonium in NRQCD\scite{Bodwin:1994jh},
\ie, bound states of heavy (mass $M$) quarks and antiquarks with typical binding momenta $1/\sqrt{MB}$ and splitting between radial excitations of order $2B$
(note the splitting between $B_2^-$ and $B_2^+$ in Table~\ref{tab.latdata_1}). The EFT developed
in Ref.~\cite{Bodwin:1994jh}~from full QCD for quarkonium also considers $\partial_t$ and $\bs{\nabla}^2/(2M)$ of the same order (compare LO \eftnopi~in Eq.~(\ref{eq.lagr}) which uses the same counting for a 
shallow nucleon state),
is approximately quark-spin independent, but retains the coupling to Coulomb and transverse gluon fields. The latter as a systematic way to incorporate the model scalar field as used in Ref.~\cite{Luke:1996hj}~from QCD could serve as an
alternative to the \eftnopi~approach if the second state is found on the lattice.
Another way to implement an excited state in an EFT considers it as an additional degree of freedom.
This leads to the so called dibaryon formulation of \eftnopi, which
introduces the $^1S_0$ and the $^3S_1$ bound state as degrees of freedom appropriately
coupled to the single-nucleon field
(application in nuclear physics: Refs.~\cite{Kaplan:1996nv,Beane:2000fi}~,
general quasi-particle mechanism: Ref.~\cite{Weinberg:1962hj}~).

Returning to the discussion of the results obtained with the EFT appropriate for the data as available at present, \ie, with single bound two-nucleon states in both channels, the analysis in
Ref.~\cite{Kirscher:2015yda}~concludes with
the calculation of Phillips and Tjon correlations. The former is between the triton binding energy $\eotri$~and the $^2S_{1/2}$ neutron-deuteron scattering length, the latter between $\eotri$~and $\eotet$. While the one-parameter correlations
exist by construction\footnote{We gratefully acknowledge comments by H.~W.~Hammer on this point.} the relatively small sensitivity \wrt~cutoff variation suggests that the above mentioned entanglement of three and four-nucleon spectra persists.
The Phillips line, in addition, was calculated down to values for $\eotri$~close to the deuteron breakup threshold. The observed divergence demonstrated the resemblance of the three and two-nucleon systems, where in the latter, the diverging
scattering length for the bound state energy approaching zero is well understood.

Addressing the breakdown of the conceived EFT with the particle number, the $^5$He and $^6$Li ground states were analyzed\scite{Barnea:2013uqa}.
$^5$He was found bound for small and unbound for larger cutoffs implying an improper
renormalization of which signatures were also found at physical \mpi~in Ref.~\cite{Kirscher:2015ana}~.
For $^6$Li, results are limited to a single cutoff value and bind
the system with the same energy per nucleon as in $\alpha$.
Both, the five and six-nucleon system as characteristic features of nuclei pose challenges to any nuclear theory: the former for its inability to sustain a bound isotope, the latter for its halo and borromean nature. To relate those
properties to either two-body ERE and triton parameters or a larger set which includes information about the short-distance structure, \eg, P-wave $NN$ phase shifts or four-nucleon resonance parameters, poses a problem
which is unresolved at physical \mpi, too. The analysis of the dependence and evolution of 5 and 6-body peculiarities on \mpi~drives the development of EFTs for larger, physical nuclei at those crucial nucleon numbers.
\section{Electromagnetism}\label{sec.enm}
While for single hadrons the effect of the U(1) gauge symmetry was calculated with the lattice methodology (\eg, the proton-neutron mass splitting\scite{Borsanyi:2014jba}),
the application to systems with $A>1$ remains impractical.
What became feasible is the analysis of the response of nuclei with $A=1,2,3,4$ to a static external electromagnetic field\scite{Chang:2015qxa}in the form of magnetic moments\scite{Beane:2014ora}and polarizabilities,
and the capture process of a neutron by a proton via the emission of a photon $np\to d\gamma$\scite{Beane:2015yha}.
In this section, we review the interplay between LQCD data and contact theories in the extraction of those observables.
\paragraph*{Data ---}
Here, as sketched above in Sec.~\ref{sec.latmet}, correlation
functions of operators with non-zero overlap with the states of 
interest provide the spectral information. In Ref.~\cite{Chang:2015qxa}~, a uniform magnetic background is 
considered with the approximation of zero-sea-quark electric charges. The approximation allows to recycle the functional determinant 
which results from the analytical integration over the quark degrees of freedom in Eq.~(\ref{eq.pathint}). It includes the U(1) 
gauge field in addition to the gluons only in the propagators for the valence quarks which stem from the contractions of the 
fields in the source and sink operators $\hat{O}$. In comparison to the spectral and scattering calculations
(first row Table~\ref{tab.lqcdpara_1}), a smaller lattice, $L^3\times T\sim (3.5~\text{fm})^3\times 5.3~$fm, but the
same~\mpill~and three mass-degenerate quark flavors (SU(3)) with a value corresponding to the physical strange-quark
mass were used.
\begin{table}
\setlength{\tabcolsep}{3.5pt}
\renewcommand{\arraystretch}{1.4}
\caption{\label{tab.latdata_2}{LQCD predictions for nuclear magnetic
 moments\scite{Beane:2014ora}$\mu$, polarizabilities\scite{Chang:2015qxa}$\beta^{(M0)}$, scattering length $a$, and 
effective range\scite{Beane:2013br}$r$ at \mpill. Magnetic moments are compared with physical
values\scite{RevModPhys.84.1527}.
Uncertainties are na\"ive sums of the combined energy shift extraction and magnetic-field-dependence fit 
uncertainties plus finite-volume effects. In the gray cells, results follow na\"ive shell-model expectations.}}
\small\centering
\begin{tabular}{l|ccccc}
\hline\hline
 State  & $a$~[fm] & $r$~[fm] & $\mu_{806}~$[nNM] & $\mu_{140}~$[nNM] &{\small $\beta^{(M0)}~[10^{-4}~\text{fm}^3]$}\\
 \hline
$n$      & $-$ & $-$                                 &\cellcolor{Grayy} $-1.981\pm 0.023$ & $-1.91$ & $1.253\pm 0.12$ \\
$p$      & $-$ & $-$                                 &\cellcolor{Grayy} $+3.119\pm 0.097$ & $+2.80$  & $5.22\pm 0.89$ \\
$nn$     & $2.33\pm 0.46$ & $1.13\pm 0.14$ & $-$                                & $-$     & $1.872\pm 0.20$ \\
$pp$     & $2.33\pm 0.46$ & $1.13\pm 0.14$ & $-$                                & $-$     & $5.31\pm 2.8$ \\
$d$      & $1.83\pm 0.31$ & $0.91\pm 0.14$ & $+1.218\pm 0.125$                  & $+0.857$ & $4.4\pm 1.8$ \\
${}^3$H  & $-$ & $-$                       &\cellcolor{Grayy} $+3.56\pm 0.23$   & $+2.98$  & $2.6\pm 1.8$ \\
${}^3$He & $-$ & $-$                       &\cellcolor{Grayy} $-2.29\pm 0.15$   & $-2.13$ & $5.4\pm 2.4$ \\
$\alpha$ & $-$ & $-$                       & $-$                                & $-$     & $3.4\pm 2.2$ \\
\hline\hline
\end{tabular}
\end{table}

The energy levels were obtained in this partially-quenched approximation from effective-mass plots as a 
function of the magnetic field strength $\vert\bs{\textbf{B}}\vert$.
By relating these levels to the expected energy eigenvalues for a\linebreak \mbox{hadron
$h\in\lbrace \text{neutron, proton, deuteron, }nn, pp, \text{${}^3$H}, \text{$^{(3,4)}$He}\rbrace$}
with charge $Q_h$, a spin $j\leq1$, $\bs{j}\sim \bs{e}_z$,
and zero momentum occupying the Landau level $n_L$:
\begin{eqnarray}\label{eq.mompolfit}
E_{h;j_z}(\bs{\textbf{B}})&=&\sqrt{M_h^2+(2n_L+1)\vert Q_he\bs{\textbf{B}}\vert}\nonumber\\
&&-\bs{\mu}_h\cdot\bs{\textbf{B}}-2\pi\beta_h^{(M0)}\vert\bs{\textbf{B}}\vert^2-2\pi\beta_h^{(M2)}\langle\hat{T}_{ij}\text{B}_i\text{B}_j\rangle+\mathcal{O}(\vert\bs{\textbf{B}}\vert^4)\;\;,
\end{eqnarray}
the magnetic moment $\bs{\mu}_h$, magnetic scalar, $\beta^{(M0)}_h$, and tensor polarizabilities
, $\beta_h^{(M2)}$, were inferred. The brackets denote the expectation value of the
traceless and symmetric combination of angular momentum generators
$\hat{j}$ $$\hat{T}_{ij}=\frac{1}{2}\left(\hat{j}_i\hat{j}_j+\hat{j}_j\hat{j}_i-\frac{2}{3}\delta_{ij}\hat{j}^2\right)\;\;.$$ 
We list the results in Table~\ref{tab.latdata_2}~and highlight the magnetic moments
of ${}^3$H and ${}^3$He as they
coincide within uncertainty margins with na\"ive expectations based on the shell model which estimates them
as a sum of $pp~(nn)$ singlet and $n~(p)$ moments as recognized in Ref.~\cite{Chang:2015qxa}~.
This behavior resembles the relations observed at physical \mpi~where the deuteron moment is the sum of
$n$ and $p$, while triton (${}^3$He) is given by the moment of $p$ ($n$), approximately. In their linear
response, the bound nuclei behave similar to external magnetic fields relative to each other.
An inference from $\mu=0$ of the $\alpha$-particle and the moments of the smaller nuclei on the more
appropriate cluster model~---~relative to the shell model for the triton and ${}^3$He~---~with ${}^3$H-$p$, ${}^3$He-$n$, and $d-d$ comprising $\alpha$
is: The $\alpha$ ground state resides predominantly in the ${}^3$He-$n$ and ${}^3$H-$p$ configurations with a slightly larger contribution
from the former.

With this data, knowledge about the response of nuclei at large \mpi~is available. The situation is peculiar enough
to be rephrased, the measurements considered the coupling of the constituents of nuclei to an external field while
disregarding the mutual interaction via the same interaction. In essence, this treatment resembles that of composite
particles in gravitational fields which takes into account the external field, only, and does not concern itself with
the way an atom, \eg, contributes to the bending of space-time. This scenario of large external fields was analyzed
in Ref.~\cite{Detmold:2015daa}~.
It was shown that for strong enough fields all two-nucleon bound states at \mpill~and
$m_\pi\sim 450~$MeV become unbound. At this unbinding field strength the scattering lengths diverge and it was
conjectured\scite{Detmold:2015daa}that at some \mpi~this field strength is the same for all $NN$ states
(see assumption in Ref.~\cite{Braaten:2003eu}~introduced in Sec.~\ref{sec.chi2nopi})~.

How the spectral data in a small magnetic background field can be utilized by a match to an EFT for the
prediction of reaction observables was also demonstrated and is summarized below.
\paragraph*{Matching to \eftnopi~---}
Under the assumption that \eftnopi~is applicable at \mpill, a generalized form of the relation Eq.~(\ref{eq.luescher})
was employed\scite{Chang:2015qxa}to relate energy eigenvalues in the presence of a background field to the LECs
which couple the magnetic field to nucleons up to next-to-leading order (NLO).
The generalization proceeds as in Ref.~\cite{Beane:2003da}~and requires the calculations of amplitude poles in a finite volume with
the interaction with the gauge field at NLO given by (dibaryon version in Ref.~\cite{Beane:2000fi}~, nucleon version in Ref.~\cite{Chen:1999tn}~)
\begin{equation}\label{eq.emlagr}
\mathcal{L}_{1\text{bdy}}=\frac{e}{2\mn}N^\dagger\left(\kappa_0+\kappa_1\tau_3\right)\bs{\sigma}\cdot\bs{\textbf{B}}N
\end{equation}
and
\begin{equation}\label{eq.emlagr}
\mathcal{L}_{2\text{bdy}}=\left[eL_1(N^TP_iN)^\dagger(N^T\overline{P}_3N)B_i-eL_2i\epsilon_{ijk}(N^TP_iN)^\dagger(N^TP_jN)B_k+\text{h.c.}\right]\;\;.
\end{equation}
The electron charge $e$, the nucleon mass $\mn$,
and the nuclear magneton of the neutron (proton) $\kappa_{n(p)}$, which defines
the isoscalar (isovector) nucleon magnetic moment
$\kappa_{0(1)}=\frac{1}{2}\left(\kappa_p\scalebox{0.7}{$\substack{+\\(-)}$}\kappa_n\right)$,
assume values as predicted by LQCD at given \mpi.
The low energy constant $L_1$ couples $^1S_0$
(projector $\overline{P}_3$) and $^3S_1$ (projector $P_i$) and thus
contributes at LO to the $np\to d\gamma$ capture.
The operator corresponding to the $L_2$ LEC does not induce transitions between spin states.
It contributes to the magnetic dipole moment of the deuteron and is thereby relevant\scite{Vanasse:2014sva} for asymmetries
in the cross sections for circularly polarized
photons impinging on an unpolarized deuteron target, $d\gamma\rightarrow np$.

In combination with the proper \eftnopi~Lagrangian governing the nuclear
interaction at NLO, the $np\to d\gamma$ amplitude
can be evaluated\scite{Chen:1999tn}. The zeros of the real part of the
inverse of this amplitude in a finite volume are
related to $k\cot\delta$\scite{Beane:2003da}.
Scattering lengths and effective ranges for the incoming singlet and outgoing
triplet were taken\scite{Chang:2015qxa}from Ref.~\cite{Beane:2013br}~to be degenerate,
$\anps\sim\anpt$ and $r_1\sim r_3$.
With the isovector $\kappa_1$ measured independently in Ref.~\cite{Chang:2015qxa}~,
$L_1$ is the only parameter left in the NLO amplitude's poles which is left
undetermined by single-nucleon and scattering parameters.
It is related to energy shifts between the singlet and triplet eigenstates in the presence of the background field\scite{Detmold:2004qn}.

The predictive power was demonstrated by calculating the cross
section of the radiative capture with the extracted value
of $L_1$ at \mpill in Ref.~\cite{Chang:2015qxa}~and \mbox{$m_\pi\sim450$~MeV in Ref.~\cite{Beane:2015yha}~}
where only the value for 806~MeV is quoted:
\begin{equation}\label{eq.cross}
\sigma_{806}(np\to d\gamma)=17~\scalebox{0.9}{$+101\choose -16$}~\text{mb}\;\;.
\end{equation}
The asymmetric uncertainty is due to the non-linear input
dependence of the cross-section which, in contrast to the renormalization of $L_1$,
uses different effective-range parameters in the singlet and triplet channel.

It is noteworthy that the LQCD results of the magnetic moments indicate a similar internal structure
of the $A=3$ nuclei as found in nature, \ie, a bound singlet with a single nucleon which determines the spin.
Relating this fact~---~the three-body response being given by that of a single nucleon~---~to the ratio of the
separation energy to the binding energy of the core singlet, at \mpill~this ratio is $\sim 2.1$ compared to
$\sim 2.9$ at physical \mpi~(see Table~\ref{tab.latdata_2}), the structure is expected to change
significantly when decreasing the pion mass where the respective ratios at 510~MeV and 300~MeV suggest a
shallow triton.
\section{Summary}\label{sec.sum}
The status of a unified description of particle and nuclear physics was presented.
This description comprises a chain of effective field theories
with QCD on the particle, \eftnopi~on the nuclear end,
and a bridge through \ce.

Interest in the sensitivity of nuclear observables to variations in fundamental
parameters~---~the pion-mass, in particular~---~arose with the inability
to solve QCD with the physical pion mass.
EFT expansions around $m_\pi=0$ were employed to assess how sensitive nuclei
react on a variation of \mpi~up to $\sim 200~$MeV.
Those attempts were reviewed in Secs.~\ref{sec.lqcd2chi} and~\ref{sec.chi2nopi}.
The introduced framework of matching a chiral EFT to data at some \mpi, and to
extrapolate nuclear amplitudes to physical or larger \mpi~while tracing carefully
implicit and explicit \mpi~dependences, is presented as an ansatz constituting one link in
the EFT chain which can be implemented once experimental data can be replaced by QCD amplitudes.
Work related to the connection between \ce~and \eftnopi, as candidate for a few-nucleon EFT, was presented in
Sec.~\ref{sec.chi2nopi}.

The following Secs.~\ref{sec.lqcdmeth} through~\ref{sec.enm} elaborate
on the approximation of QCD in light nuclei by
matching contact EFTs to lattice calculations.
The methodology of LQCD is overviewed with a focus on uncertainties presumably responsible for inconsistent
EFT postdictions. The available nuclear LQCD data is compiled in
Tables~\ref{tab.latdata_1} and~\ref{tab.latdata_2},
including only results obtained with controlled approximations.

In Sec.~\ref{sec.hal} and~\ref{sec.left}, we introduced two methods of deriving a nuclear effective
interaction. A method based on a velocity expansion of a potential consistent with a QCD wave function, and
the adaption of an \eftnopi~analog to LQCD amplitudes.
The latter method is included in the summary of efforts to assess the electromagnetic structure of nuclei
in Sec.~\ref{sec.enm}~.

\section{Outlook}\label{sec.out}
It is one aim of this article to review research on the extrapolation of QCD solved at unphysical pion
masses to physical pion masses \textit{as well as} the consequences of an enlarged pion mass on nuclear systems.
Future work as suggested below concerns both aspects.
\paragraph{Outlook in nucleon number ---}
The anecdotal introductory hint to the resemblance between the contemporary numerical effort
and the historical, experimental one does not apply to larger nuclei. Their properties will
not be accessible with LQCD within a similar time frame as physical experiments ventured beyond the deuteron.
As cluster/halo EFTs\footnote{general frame work in Ref.~\cite{Bertulani:2002sz}~;
treatment of narrow resonances in Refs.~\cite{Bedaque:2003wa,Gelman:2009be}~;
applications: \mbox{proton halo\scite{Ryberg:2013iga}}, \mbox{$\alpha\alpha$ in Ref.~\cite{Higa:2008dn}~}, \mbox{two-neutron halos\scite{Canham:2008jd}},
\mbox{${}^{7}\text{Li}+n\to{}^{8}\text{Li}+\gamma$ in Ref.~\cite{Zhang:2013kja}~},
\mbox{${}^{7}\text{Be}+p\to{}^{8}\text{B}+\gamma$ in Ref.~\cite{Zhang:2015vew}~}, \mbox{$d+t\to N+\alpha$ in Ref.~\cite{Brown:2013zla}~}.}
constitute the next link in the EFT chain relating QCD parameters to
nuclei with $A>4$ and have so far been renormalized through a match to
data\footnote{The work in Ref.~\cite{Zhang:2013kja}~can be considered the first rigorous matching of a microscopic and a cluster EFT.},
unavailable from LQCD, their connection to \eftnopi~is crucial. Whether or not the bulk properties of $A>4$ nuclei are a universal
consequence of two- and three-nucleon data and thus could already be parameterized with \mpi~is unknown.
Specifically, do the unbound Helium-5, the shallow $\alpha N$ resonances, the halo structure of Helium-6
emerge at LO \eftnopi? It is an open question if an additional renormalization condition in form of a five or six-body counter term are necessary to put those poles in the respective LO amplitudes, or
if those observables are sensitive to two-body P-wave interactions as which they should not be
considered before next-to-next-to-leading order.
Once EFT and few-body practitioners have addressed this question at physical \mpi, the amplitudes
can be matched to cluster EFTs and thereby pass the \mpi~dependence to larger nuclei, systematically.
\paragraph{Refining the interaction ---}
The role of the electromagnetic interaction between quarks for lattice nuclei is unknown but
the response of nuclei to external magnetic fields \textit{has} been explored in LQCD measurements.
The latter included a fascinating demonstration how this method can probe extreme conditions inaccessible by
experiments. These analyses were covered in Sec.~\ref{sec.enm}~.
Next to the numerical effort to implement the electromagnetic interaction
in LQCD calculations, there remain conceptual issues hampering\footnote{For recent progress in the systematic treatment of the
Coulomb force see Refs.~\cite{Konig:2015aka,Kirscher:2015zoa,Vanasse:2014kxa}~.}
their \eftnopi~consideration.
This is an instance where EFT can make predictions by including the long-range Coulomb force in
\eftnopi~at large pion masses.
Contrary to nature, the proton-proton system provides both a bound and scattering specimen to
study the effect resulting from the combination of a long and a presumably relatively
short-ranged force.
At present, we assume
that the energy gap between the triton and Helium-3 is approximately an invariant \wrt~changes in \mpi.
Noting a peculiar consequence of
a widening gap, namely a conceivable unbound helion in the presence of a shallowly bound triton,
shall motivate work in that direction.

Of interest for understanding the difference between hyper and ordinary nuclei are LQCD measurements at a fixed pion mass for both, the SU(3) symmetric point,
and with a shifted strange-quark mass $m_s$, \ie, explicit breaking of the flavor symmetry. The effect of, \eg, an infinitesimal shift in $m_s$ on the two states of ${}^3_\Lambda$H observed at \mpill~by NPLQCD,
could indicate the significance of the SU(3) breaking relative to that of the pion mass for the shallowness of the hyper triton \wrt~to the ordinary triton.
The LQCD data available on strange nuclei has not yet been matched to a contact theory at 450~or 806~MeV. This requires a generalization of the SU(2) isospin-symmetric Lagrangian of \eftnopi~to
SU(3)\footnote{We acknowledge the explanation by M.~Elyahu and N.~Barnea to whom this idea belongs.}
and thus a comprehensive theory for the baryon octet. In combination with LQCD data, which in contrast to nature is roughly as accurate for the
$s=0$ as it is for the $s=-1$ sector, will allow for a systematic study of the peculiar differences between strange and ordinary nuclei.

The observed \mpi-insensitivity of the approximate SU(4) symmetry of the nuclear interaction as shown in Fig.~\ref{fig.lecs}
allows for the investigation of the relevant QCD parameters which cause this remarkable feature. This sensitivity analysis can be
performed at large pion masses since \mpi~does not seem to be significant for the effect. We are thus
able to understand the mechanism behind the breaking of Wigner's SU(4) symmetry also at physical \mpi~---~an insight that would
demonstrate the beneficial interplay between LQCD and nuclear EFTs.
\paragraph*{Understanding \mpi~sensitivity ---}
To predict low-energy renormalization-group fixed points of QCD, we desire data about
the smooth dependence of nuclear spectra from LQCD.
Available lattice calculations investigate nuclei at isolated pion masses. For the interpolation between them,
no theoretical ansatz is known for \mpi~beyond the convergence rate of ChPT. Even within that radius,
the murky formulation of \ce~in the few-nucleon sector hampers the uncertainty quantification of those interpolations and
renders them less useful.
To develop or refine interpolating and extrapolating theories,
knowledge of whether or not a critical RG flow trajectory is approached will be of use.
LQCD analyses of energy gaps between $nn$ singlet and the deuteron, deuteron and
triton, triton and $\alpha$, and $\alpha$ and Helium-5 at two infinitesimally close \mpi's would indicate whether or not one approaches a critical \mpi~for
any of those systems.

Such an analysis touches the question how parameters of constituents characterize compounds, \eg, like the infinite two-nucleon scattering lengths in combination with a shallow
triton ``furnish'' the $\alpha$-particle.
The aforementioned cluster EFTs work well due to a separation of scales between the excitation energy of the $\alpha$ and the shallow $\alpha N$ poles.
The ratio between the binding energies of Helium-5~---~if bound at all~---~and the $\alpha$ as one characteristic of nuclei at physical \mpi,
namely their shell structure, is one crucial observable which as a ratio is more accessible to LQCD than the bare values. To analyze the \mpi~dependence of it
would shed light on the emergence of the shell structure, the peculiar mass gap at $A=5$, and the prominent role of the $\alpha$ as a building block
for larger nuclei.

How nuclear \textit{two}-body systems can exhibit a peculiar behavior like a Feshbach resonance has been shown by simulating extreme magnetic fields
which reside naturally only in cosmological objects like magnetars. Similar features of larger systems, like the development of a four-nucleon Efimov spectrum due
to a triton near the deuteron threshold~---~a scenario which is admissible within error bars at $m_\pi=300~$MeV~---~and thus a mass gap at $A=3$ is of
undeniable empirical value to identify the underlying QCD mechanisms for such characteristics of the nuclear chart. To that end, ratios of binding energies are
more important for the theoretical understanding than relatively less accurately measurable absolute binding energies.
\paragraph*{Focal-point system ---}
The five-baryon system as a gateway to heavier nuclei and refined EFTs concludes this article. \textit{First}, it poses a challenge for numerical techniques, for LQCD as well as traditional few-body methods.
\textit{Second}, because its features are neither understood as emergent or unique. Restated, whether or not conventional \eftnopi~applies to it is unknown. \textit{Third},
its amplitudes are the canonical candidates for the bridge between single-baryon and cluster EFTs. The pion-mass dependence of the dynamics of this system is
therefore key to an understanding of the emergence of complex phenomena in nuclei from the interactions governing its basic building blocks.
\section*{Acknowledgements}
The author gratefully acknowledges the hospitality of the University
of Trento and F.~Pederiva, the participants of the ECT$^*$ workshop on lattice nuclei,
in particular conversations with S.~Aoki, N.~Barnea, D.~Gazit, U.~van Kolck, and M.~Savage, comments
on the manuscript by E.~Epelbaum, and the financial support of the Minerva Foundation.
\bibliographystyle{ws-ijmpe}
\bibliography{refs_r2}
\end{document}